\documentclass[longauth]{aa}
\usepackage{graphicx}
\usepackage{gensymb}
\usepackage{txfonts}
\usepackage[flushleft]{threeparttable}
\renewcommand{\arraystretch}{1.1}
\usepackage[hidelinks,colorlinks=true,linkcolor=blue,citecolor=blue]{hyperref}
\usepackage{longtable,tabu}
\usepackage{lipsum}
\usepackage{adjustbox}
\usepackage{stfloats}
\usepackage{scrextend}
\usepackage{subcaption}
\usepackage{amssymb}
\usepackage{caption}
\usepackage{amsmath}
\usepackage{amssymb}
\usepackage{mathrsfs}
\usepackage{multicol}
\usepackage{supertabular}
\usepackage{orcidlink}
\usepackage{tabularx}
\usepackage{float}
\usepackage[dvipsnames]{xcolor}
\newcommand{\feh}{\mbox{[Fe/H]}\xspace}
\newcommand{\sih}{\mbox{[Si/H]}\xspace}
\newcommand{\mgh}{\mbox{[Mg/H]}\xspace}
\newcommand{\teff}{\mbox{$\it{T_{\rm \mathrm{eff}}}$}\xspace}

\newcommand{\ms}{\mbox{m\,s$^{-1}$}\xspace}

\newcommand{\mplanet}{\mbox{$M_{\rm P}$}\xspace}
\newcommand{\rplanet}{\mbox{$R_{\rm P}$}\xspace}
\newcommand{\mjup}{\mbox{$\it{M_{\rm \mathrm{Jup}}}$}\xspace}

\newcommand{\me}{\mbox{$\it{M_{\rm \mathrm{\oplus}}}$}\xspace}
\newcommand{\re}{\mbox{$\it{R_{\rm \mathrm{\oplus}}}$}\xspace}

\newcommand{\rstar}{\mbox{$R_{*}$}\xspace}

\newcommand{\msol}{\mbox{$\it{M_\mathrm{\odot}}$}\xspace}
\newcommand{\rsol}{\mbox{$\it{R_\mathrm{\odot}}$}\xspace}
\newcommand{\se}{\mbox{$\it{S_{\rm \mathrm{\oplus}}}$}\xspace}

\newcommand{\denssol}{\mbox{$\it{\rho_\mathrm{\odot}}$}\xspace}

\newcommand{\CONAN}{{\sc \tt CONAN}\xspace}
\newcommand{\plaNETic}{{\sc \tt plaNETic}\xspace}
\newcommand{\ARDENT}{{\sc \tt ARDENT}\xspace}

\newcommand{\ldtk}{{\sc \tt LDTK}\xspace}

\AtBeginDocument{}
\usepackage{xcolor}

\begin{document}

\title{TOI-1699A: a sub-Neptune and a super-Jovian in an S-type planetary system}

\authorrunning{A. Psaridi et al.}

\author{
Angelica~Psaridi$^{\star}$\orcidlink{0000-0002-4797-2419}\inst{\ref{inst:1},\ref{inst:2}}
\and
Manu~Stalport\orcidlink{0000-0003-0996-6402}\inst{\ref{inst:3},\ref{inst:4}}
\and
Giacomo~Mantovan\orcidlink{0000-0002-6871-6131}\inst{\ref{inst:5},\ref{inst:6}}
\and
Arianna~Nigioni\orcidlink{0009-0004-5882-6574}\inst{\ref{inst:7}}
\and
Julia~Venturini\orcidlink{0000-0001-9527-2903}\inst{\ref{inst:7}}
\and
Andrea~Fortier\orcidlink{0000-0001-8450-3374}\inst{\ref{inst:8},\ref{inst:9}}
\and
Davide~Gandolfi\orcidlink{0000-0001-8627-9628}\inst{\ref{inst:10}}
\and
Alexis~Heitzmann\orcidlink{0000-0002-8091-7526}\inst{\ref{inst:7}}
\and
Solène~Ulmer-Moll\orcidlink{0000-0003-2417-7006}\inst{\ref{inst:3},\ref{inst:11}}
\and
Andrea~Bonfanti\orcidlink{0000-0002-1916-5935}\inst{\ref{inst:12}}
\and
Jo~Ann~Egger\orcidlink{0000-0003-1628-4231}\inst{\ref{inst:13}}
\and
Sérgio~G.~Sousa\orcidlink{0000-0001-9047-2965}\inst{\ref{inst:14},\ref{inst:15}}
\and
Thomas~G.~Wilson\orcidlink{0000-0001-8749-1962}\inst{\ref{inst:16}}
\and
Gaia~Lacedelli\orcidlink{0000-0002-4197-7374}\inst{\ref{inst:17},\ref{inst:18}}
\and
Emma~Nabbie\orcidlink{0000-0003-0571-2245}\inst{\ref{inst:19}}
\and
Karen~A.~Collins\orcidlink{0000-0001-6588-9574}\inst{\ref{inst:19}}
\and
Alexandre~C.~M.~Correia\orcidlink{0000-0002-8946-8579}\inst{\ref{inst:20}}
\and
David~Degen\orcidlink{0009-0008-1068-481X}\inst{\ref{inst:21}}
\and
Alexander~J.~Mustill\orcidlink{0000-0002-2086-3642}\inst{\ref{inst:22}}
\and
Vardan~Adibekyan\orcidlink{0000-0002-0601-6199}\inst{\ref{inst:14},\ref{inst:15}}
\and
Elise~Furlan\orcidlink{0000-0001-9800-6248}\inst{\ref{inst:23}}
\and
Alexis~Brandeker\orcidlink{0000-0002-7201-7536}\inst{\ref{inst:24}}
\and
Jorge~Lillo-Box\orcidlink{0000-0003-3742-1987}\inst{\ref{inst:25}}
\and
Felipe~Murgas\orcidlink{0000-0001-9087-1245}\inst{\ref{inst:17},\ref{inst:18}}
\and
Nuno~C.~Santos\orcidlink{0000-0003-4422-2919}\inst{\ref{inst:14},\ref{inst:15}}
\and
Amadeo~Castro~González\orcidlink{0000-0001-7439-3618}\inst{\ref{inst:7}}
\and
Babatunde~Akinsanmi\orcidlink{0000-0001-6519-1598}\inst{\ref{inst:7},\ref{inst:aus}}
\and
Yann~Alibert\orcidlink{0000-0002-4644-8818}\inst{\ref{inst:8},\ref{inst:9}}
\and
Roi~Alonso\orcidlink{0000-0001-8462-8126}\inst{\ref{inst:17},\ref{inst:18}}
\and
Tamás~Bárczy\orcidlink{0000-0002-7822-4413}\inst{\ref{inst:26}}
\and
David~Barrado\orcidlink{0000-0002-5971-9242}\inst{\ref{inst:25}}
\and
Susana~C.~C.~Barros\orcidlink{0000-0003-2434-3625}\inst{\ref{inst:14},\ref{inst:15}}
\and
Wolfgang~Baumjohann\orcidlink{0000-0001-6271-0110}\inst{\ref{inst:12}}
\and
Willy~Benz\orcidlink{0000-0001-7896-6479}\inst{\ref{inst:8},\ref{inst:9}}
\and
Nicolas~Billot\orcidlink{0000-0003-3429-3836}\inst{\ref{inst:7}}
\and
Federico~Biondi\orcidlink{0000-0002-1337-3653}\inst{\ref{inst:6},\ref{inst:27}}
\and
Luca~Borsato\orcidlink{0000-0003-0066-9268}\inst{\ref{inst:6}}
\and
Christopher~Broeg\orcidlink{0000-0001-5132-2614}\inst{\ref{inst:8},\ref{inst:9}}
\and
Andrew~Collier~Cameron\orcidlink{0000-0002-8863-7828}\inst{\ref{inst:28}}
\and
Szilárd~Csizmadia\orcidlink{0000-0001-6803-9698}\inst{\ref{inst:29}}
\and
Patricio~E.~Cubillos\orcidlink{0000-0002-1347-2600}\inst{\ref{inst:12},\ref{inst:30}}
\and
Melvyn~B.~Davies\orcidlink{0000-0001-6080-1190}\inst{\ref{inst:31}}
\and
Jerome~de~Leon\orcidlink{0000-0002-6424-3410}\inst{\ref{inst:32}}
\and
Magali~Deleuil\orcidlink{0000-0001-6036-0225}\inst{\ref{inst:33}}
\and
Adrien~Deline\inst{\ref{inst:7}}
\and
Olivier~D.~S.~Demangeon\orcidlink{0000-0001-7918-0355}\inst{\ref{inst:14},\ref{inst:15}}
\and
Brice-Olivier~Demory\orcidlink{0000-0002-9355-5165}\inst{\ref{inst:8},\ref{inst:9},\ref{inst:34}}
\and
Aliz~Derekas\orcidlink{0000-0002-6526-9444}\inst{\ref{inst:35}}
\and
Billy~Edwards\orcidlink{0000-0002-5494-3237}\inst{\ref{inst:36}}
\and
David~Ehrenreich\orcidlink{0000-0001-9704-5405}\inst{\ref{inst:7},\ref{inst:37}}
\and
Anders~Erikson\inst{\ref{inst:29}}
\and
Yoshi~N.~E.~Eschen\orcidlink{0009-0006-6397-2503}\inst{\ref{inst:16}}
\and
Luca~Fossati\orcidlink{0000-0003-4426-9530}\inst{\ref{inst:12}}
\and
Malcolm~Fridlund\orcidlink{0000-0002-0855-8426}\inst{\ref{inst:11},\ref{inst:38}}
\and
Izuru~Fukuda\orcidlink{0000-0002-9436-2891}\inst{\ref{inst:39}}
\and
Akihiko~Fukui\orcidlink{0000-0002-4909-5763}\inst{\ref{inst:17},\ref{inst:32}}
\and
Tianjun~Gan\orcidlink{0000-0002-4503-9705}\inst{\ref{inst:17},\ref{inst:18}}
\and
Kosmas~Gazeas\orcidlink{0000-0002-8855-3923}\inst{\ref{inst:40}}
\and
Michaël~Gillon\orcidlink{0000-0003-1462-7739}\inst{\ref{inst:4}}
\and
Manuel~Güdel\orcidlink{0000-0001-9818-0588}\inst{\ref{inst:41}}
\and
Maximilian~N.~Günther\orcidlink{0000-0002-3164-9086}\inst{\ref{inst:13}}
\and
Yuya~Hayashi\orcidlink{0000-0001-8877-0242}\inst{\ref{inst:39}}
\and
Christiane~Helling\orcidlink{0000-0002-8275-1371}\inst{\ref{inst:12},\ref{inst:42}}
\and
Kate~G.~Isaak\orcidlink{0000-0001-8585-1717}\inst{\ref{inst:13}}
\and
László~L.~Kiss\orcidlink{0000-0002-3234-1374}\inst{\ref{inst:43},\ref{inst:44}}
\and
Daniel~Kitzmann\orcidlink{0000-0003-4269-3311}\inst{\ref{inst:8},\ref{inst:9}}
\and
Emanuel~Kopp\inst{\ref{inst:29}}
\and
Judith~Korth\orcidlink{0000-0002-0076-6239}\inst{\ref{inst:17},\ref{inst:18}}
\and
Kristine~W.~F.~Lam\orcidlink{0000-0002-9910-6088}\inst{\ref{inst:29}}
\and
Jacques~Laskar\orcidlink{0000-0003-2634-789X}\inst{\ref{inst:45}}
\and
Alain~Lecavelier~des~Etangs\orcidlink{0000-0002-5637-5253}\inst{\ref{inst:46}}
\and
Adrien~Leleu\orcidlink{0000-0003-2051-7974}\inst{\ref{inst:7},\ref{inst:8}}
\and
Monika~Lendl\orcidlink{0000-0001-9699-1459}\inst{\ref{inst:7}}
\and
Kathryn~V.~Lester\inst{\ref{inst:47}}
\and
Armin~Luntzer\orcidlink{0000-0002-9449-9919}\inst{\ref{inst:41}}
\and
Demetrio~Magrin\orcidlink{0000-0003-0312-313X}\inst{\ref{inst:6}}
\and
Pierre~F.~L.~Maxted\orcidlink{0000-0003-3794-1317}\inst{\ref{inst:48}}
\and
Bruno~Merín\orcidlink{0000-0002-8555-3012}\inst{\ref{inst:49}}
\and
Christoph~Mordasini\orcidlink{0000-0002-1013-2811}\inst{\ref{inst:8},\ref{inst:9}}
\and
Norio~Narita\orcidlink{0000-0001-8511-2981}\inst{\ref{inst:17},\ref{inst:32},\ref{inst:50}}
\and
Valerio~Nascimbeni\orcidlink{0000-0001-9770-1214}\inst{\ref{inst:6}}
\and
Göran~Olofsson\orcidlink{0000-0003-3747-7120}\inst{\ref{inst:24}}
\and
Hugh~P.~Osborn\orcidlink{0000-0002-4047-4724}\inst{\ref{inst:8}}
\and
Roland~Ottensamer\orcidlink{0000-0001-5684-5836}\inst{\ref{inst:41}}
\and
Isabella~Pagano\orcidlink{0000-0001-9573-4928}\inst{\ref{inst:51}}
\and
Enric~Pallé\orcidlink{0000-0003-0987-1593}\inst{\ref{inst:17},\ref{inst:18}}
\and
L\'ena~Parc\orcidlink{0000-0002-7382-1913}\inst{\ref{inst:7}}
\and
Hannu~Parviainen\orcidlink{0000-0001-5519-1391}\inst{\ref{inst:17},\ref{inst:18}}
\and
Austin~Peel\orcidlink{0000-0003-0488-8978}\inst{\ref{inst:8},\ref{inst:9}}
\and
Gisbert~Peter\orcidlink{0000-0001-6101-2513}\inst{\ref{inst:29}}
\and
Daniele~Piazza\inst{\ref{inst:8}}
\and
Giampaolo~Piotto\orcidlink{0000-0002-9937-6387}\inst{\ref{inst:6},\ref{inst:52}}
\and
Don~Pollacco\orcidlink{0000-0001-9850-9697}\inst{\ref{inst:16}}
\and
Didier~Queloz\orcidlink{0000-0002-3012-0316}\inst{\ref{inst:21},\ref{inst:53}}
\and
Roberto~Ragazzoni\orcidlink{0000-0002-7697-5555}\inst{\ref{inst:6},\ref{inst:52}}
\and
Nicola~Rando\inst{\ref{inst:13}}
\and
Francesco~Ratti\inst{\ref{inst:13}}
\and
Heike~Rauer\orcidlink{0000-0002-6510-1828}\inst{\ref{inst:54},\ref{inst:55}}
\and
Ignasi~Ribas\orcidlink{0000-0002-6689-0312}\inst{\ref{inst:1},\ref{inst:2}}
\and
Gaetano~Scandariato\orcidlink{0000-0003-2029-0626}\inst{\ref{inst:51}}
\and
Richard~P.~Schwarz\orcidlink{0000-0001-8227-1020}\inst{\ref{inst:19}}
\and
Nicholas~J.~Scott\orcidlink{0000-0003-1038-9702}\inst{\ref{inst:56}}
\and
Damien~Ségransan\orcidlink{0000-0003-2355-8034}\inst{\ref{inst:7}}
\and
Attila~E.~Simon\orcidlink{0000-0001-9773-2600}\inst{\ref{inst:8},\ref{inst:9}}
\and
Alexis~M.~S.~Smith\orcidlink{0000-0002-2386-4341}\inst{\ref{inst:29}}
\and
Toshi~Suganuma\orcidlink{0009-0007-2926-1924}\inst{\ref{inst:39}}
\and
Sophia~Sulis\orcidlink{0000-0001-8783-526X}\inst{\ref{inst:33}}
\and
Gyula~M.~Szabó\orcidlink{0000-0002-0606-7930}\inst{\ref{inst:35}}
\and
Benjamin~M.~Tofflemire\orcidlink{0000-0003-2053-0749}\inst{\ref{inst:57},\ref{inst:58}}
\and
Stéphane~Udry\orcidlink{0000-0001-7576-6236}\inst{\ref{inst:7}}
\and
Valérie~Van~Grootel\orcidlink{0000-0003-2144-4316}\inst{\ref{inst:3}}
\and
Eva~Villaver\orcidlink{0000-0003-4936-9418}\inst{\ref{inst:17},\ref{inst:18}}
\and
Hugo~G.~Vivien\orcidlink{0000-0001-7239-6700}\inst{\ref{inst:33}}
\and
Nicholas~A.~Walton\orcidlink{0000-0003-3983-8778}\inst{\ref{inst:59}}
\and
Sebastian~Wolf\inst{\ref{inst:8}}
}

\institute{
\label{inst:1} Institut de Ci\`encies de l'Espai (ICE, CSIC), Campus UAB, Can Magrans s/n, 08193 Bellaterra, Spain \and
\label{inst:2} Institut d'Estudis Espacials de Catalunya (IEEC), 08860 Castelldefels, Barcelona, Spain \and
\label{inst:3} Space sciences, Technologies and Astrophysics Research (STAR) Institute, Universit\'e de Li\`ege, All\'ee du 6 Ao\^ut 19C, 4000 Li\`ege, Belgium \and
\label{inst:4} Astrobiology Research Unit, Universit\'e de Li\`ege, All\'ee du 6 Ao\^ut 19C, 4000 Li\`ege, Belgium \and
\label{inst:5} Centro di Ateneo di Studi e Attivit\`a Spaziali G. Colombo -- Universit\`a degli Studi di Padova, Via Venezia 15, 35131 Padova, Italy \and
\label{inst:6} INAF, Osservatorio Astronomico di Padova, Vicolo dell'Osservatorio 5, 35122 Padova, Italy \and
\label{inst:7} Observatoire astronomique de l'Universit\'e de Gen\`eve, Chemin Pegasi 51, 1290 Versoix, Switzerland \and
\label{inst:8} Space Research and Planetary Sciences, Physics Institute, University of Bern, Gesellschaftsstrasse 6, 3012 Bern, Switzerland \and
\label{inst:9} Center for Space and Habitability, University of Bern, Gesellschaftsstrasse 6, 3012 Bern, Switzerland \and
\label{inst:10} Dipartimento di Fisica, Universit\`a degli Studi di Torino, via Pietro Giuria 1, 10125 Torino, Italy \and
\label{inst:11} Leiden Observatory, University of Leiden, Einsteinweg 55, 2333 CA Leiden, The Netherlands \and
\label{inst:12} Space Research Institute, Austrian Academy of Sciences, Schmiedlstrasse 6, 8042 Graz, Austria \and
\label{inst:13} European Space Agency (ESA), European Space Research and Technology Centre (ESTEC), Keplerlaan 1, 2201 AZ Noordwijk, The Netherlands \and
\label{inst:14} Instituto de Astrofísica e Ciências do Espaço, Universidade do Porto, CAUP, Rua das Estrelas, 4150-762 Porto, Portugal \and
\label{inst:15} Departamento de F\'isica e Astronomia, Faculdade de Ci\^encias, Universidade do Porto, Rua do Campo Alegre, 4169-007 Porto, Portugal \and
\label{inst:16} Department of Physics, University of Warwick, Gibbet Hill Road, Coventry CV4 7AL, United Kingdom \and
\label{inst:17} Instituto de Astrof\'isica de Canarias, V\'ia L\'actea s/n, 38200 La Laguna, Tenerife, Spain \and
\label{inst:18} Departamento de Astrof\'isica, Universidad de La Laguna (ULL), 38206 La Laguna, Tenerife, Spain \and
\label{inst:19} Center for Astrophysics \textbar\ Harvard \& Smithsonian, 60 Garden Street, Cambridge, MA 02138, USA \and
\label{inst:20} CFisUC, Departamento de F\'isica, Universidade de Coimbra, 3004-516 Coimbra, Portugal \and
\label{inst:21} ETH Zurich, Department of Physics, Wolfgang-Pauli-Strasse 2, 8093 Zurich, Switzerland \and
\label{inst:22} Lund Observatory, Division of Astrophysics, Department of Physics, Lund University, Box 118, 221 00 Lund, Sweden \and
\label{inst:23} NASA Exoplanet Science Institute, IPAC, California Institute of Technology, Pasadena, CA 91125, USA \and
\label{inst:24} Department of Astronomy, Stockholm University, AlbaNova University Center, 10691 Stockholm, Sweden \and
\label{inst:25} Centro de Astrobiolog\'ia (CSIC-INTA), ESAC campus, 28692 Villanueva de la Ca\~nada, Madrid, Spain \and
\label{inst:aus} School of Physics, University of New South Wales, Kensington, NSW 2052, Australia \and
\label{inst:26} Admatis, 5. Kand\'o K\'alm\'an Street, 3534 Miskolc, Hungary \and
\label{inst:27} Max-Planck-Institut f\"ur extraterrestrische Physik, Gie{\ss}enbachstra{\ss}e 1, 85748 Garching, Germany \and
\label{inst:28} Centre for Exoplanet Science, SUPA School of Physics and Astronomy, University of St Andrews, North Haugh, St Andrews KY16 9SS, United Kingdom\and
\label{inst:29} Institut f\"ur Weltraumforschung, Deutsches Zentrum f\"ur Luft- und Raumfahrt (DLR), Rutherfordstra\ss e~2, 12489 Berlin, Germany \and
\label{inst:30} INAF, Osservatorio Astrofisico di Torino, Via Osservatorio, 20, 10025 Pino Torinese, Italy \and
\label{inst:31} Centre for Mathematical Sciences, Lund University, Box 118, 221 00 Lund, Sweden \and
\label{inst:32} Komaba Institute for Science, The University of Tokyo, 3-8-1 Komaba, Meguro, Tokyo 153-8902, Japan \and
\label{inst:33} Aix Marseille Université, CNRS, CNES, Institut Origines, LAM, Marseille, France \and
\label{inst:34} Artificial Organs (ARTORG) Center for Biomedical Engineering Research, University of Bern, Bern, Switzerland \and
\label{inst:35} E\"otv\"os Lor\'and University, Gothard Astrophysical Observatory, 9700 Szombathely, Szent Imre h. u. 112, Hungary \and
\label{inst:36} Stichting Ruimteonderzoek Nederland (SRON), Netherlands Institute for Space Research, Niels Bohrweg 4, 2333 CA Leiden, The Netherlands \and
\label{inst:37} Centre Vie dans l'Univers, Facult\'e des sciences, Universit\'e de Gen\`eve, Quai Ernest-Ansermet 30, 1211 Gen\`eve 4, Switzerland \and
\label{inst:38} Department of Space, Earth and Environment, Chalmers University of Technology, Onsala Space Observatory, 439 92 Onsala, Sweden \and
\label{inst:39} Department of Multi-Disciplinary Sciences, Graduate School of Arts and Sciences, The University of Tokyo, 3-8-1 Komaba, Meguro, Tokyo 153-8902, Japan \and
\label{inst:40} National and Kapodistrian University of Athens, Department of Physics, University Campus, Zografou, 157 84 Athens, Greece \and
\label{inst:41} Department of Astrophysics, University of Vienna, T\"urkenschanzstrasse 17, 1180 Vienna, Austria \and
\label{inst:42} Institute for Theoretical Physics and Computational Physics, Graz University of Technology, Petersgasse 16, 8010 Graz, Austria \and
\label{inst:43} Konkoly Observatory, Research Centre for Astronomy and Earth Sciences, 1121 Budapest, Konkoly Thege Mikl\'os \'ut 15-17, Hungary \and
\label{inst:44} E\"otv\"os Lor\'and University, Institute of Physics, P\'azm\'any P\'eter s\'et\'any 1/A, 1117 Budapest, Hungary \and
\label{inst:45} Institut de M\'ecanique C\'eleste et de Calcul des \'Eph\'em\'erides (IMCCE), UMR8028 CNRS, Observatoire de Paris, PSL Univ., Sorbonne Univ., 77 av. Denfert-Rochereau, 75014 Paris, France \and
\label{inst:46} Institut d'astrophysique de Paris, UMR7095 CNRS, Universit\'e Pierre \& Marie Curie, 98bis blvd. Arago, 75014 Paris, France \and
\label{inst:47} Mount Holyoke College, South Hadley, MA 01075, USA \and
\label{inst:48} Astrophysics Group, Lennard Jones Building, Keele University, Staffordshire, ST5 5BG, United Kingdom \and
\label{inst:49} European Space Agency (ESA), European Space Astronomy Centre (ESAC), Camino Bajo del Castillo s/n, 28692 Villanueva de la Ca\~nada, Madrid, Spain \and
\label{inst:50} Astrobiology Center, 2-21-1 Osawa, Mitaka, Tokyo 181-8588, Japan \and
\label{inst:51} INAF, Osservatorio Astrofisico di Catania, Via S. Sofia 78, 95123 Catania, Italy \and
\label{inst:52} Dipartimento di Fisica e Astronomia "Galileo Galilei", Universit\`a degli Studi di Padova, Vicolo dell'Osservatorio 3, 35122 Padova, Italy \and
\label{inst:53} Cavendish Laboratory, JJ Thomson Avenue, Cambridge CB3 0HE, United Kingdom\and
\label{inst:54} Deutsches Zentrum f\"ur Luft- und Raumfahrt (DLR), Markgrafenstra{\ss}e 37, 10117 Berlin, Germany \and
\label{inst:55} Institut f\"ur Geologische Wissenschaften, Freie Universit\"at Berlin, Malteserstra{\ss}e 74-100, 12249 Berlin, Germany \and
\label{inst:56} The CHARA Array of Georgia State University, Mount Wilson Observatory, Mount Wilson, CA 91023, USA \and
\label{inst:57} Search for Extraterrestrial Intelligence (SETI) Institute, Mountain View, CA 94043, USA \and
\label{inst:58} NASA Ames Research Center, Moffett Field, CA 94035, USA \and
\label{inst:59} Institute of Astronomy, University of Cambridge, Madingley Road, Cambridge, CB3 0HA, United Kingdom
}

\date{Received 19 September 2026 / Accepted 10 November 2026}

    \abstract{We present the detection and characterization of the TOI-1699A (TYC 3415-00428-1) S-type planetary system, hosted by the G5\,V primary of a wide binary with an M3\,V stellar companion at a projected separation of ${\sim}348$~au. The inner planet, TOI-1699A b, was initially detected by TESS and confirmed with targeted CHEOPS photometry, ground-based photometry, and HARPS-N follow-up. With an orbital period of ${\sim}$4.55 d, radius of 3.096$^{+0.057}_{-0.056}$~\re, mass of $13.9^{+1.9}_{-1.8}$~\me, and bulk density of 2.58$^{+0.40}_{-0.39}$~g~cm$^{-3}$, TOI-1699A b lies firmly in the sub-Neptune regime and requires a substantial volatile envelope. The HARPS-N data additionally reveal a massive outer companion, TOI-1699A c, with an orbital period of 108.75 $\pm$ 0.10~days, a low but significant eccentricity of 0.0737 $\pm$ 0.0034, and a minimum mass of 5.54 $\pm$ 0.21~$M_{\mathrm{Jup}}$. Assuming coplanarity with the inner planet, TOI-1699A c is not expected to transit, and our dynamical stability analysis disfavors a stellar-mass nature and limits its true mass to $M_{\mathrm{c}} \lesssim 18~\mjup$. Our combined data-driven and dynamical detection-limit analysis excludes additional planets between 54 and 235~days, and planets more massive than 27~\me\ at shorter periods. TOI-1699A joins the still limited sample of S-type multiplanet systems, while upcoming \textit{Gaia}~DR4 astrometry together with radial velocity monitoring can constrain the true mass and inclination of the outer planet.}

   \maketitle
   \begingroup
    \renewcommand{\thefootnote}{$\star$}
    \footnotetext{Corresponding author: \texttt{apsaridi@ice.csic.es}}
    \endgroup
    \nolinenumbers
    \renewcommand{\linenumbers}{}

\section{Introduction}\label{sec:Introduction}
Sub-Neptunes, planets with radii between $\sim$2 and 4 \re, are among the most common outcomes of planet formation at short orbital periods, despite having no counterpart in the Solar System (\citealt{Petigura2013}; \citealt{Fulton2017}). Their bulk properties suffer from a compositional degeneracy: a given mass and radius can be reproduced by either a rocky interior with a low-mass H/He envelope (``gas dwarfs'') or a volatile-rich, water-dominated interior (``water worlds''; \citealt{Zeng2019}; \citealt{LuquePalle2022}; \citealt{Aguichine2025}). Distinguishing between those two scenarios and testing the mass-loss and formation mechanisms proposed to shape the radius valley (\citealt{OwenWu2017}; \citealt{Ginzburg2018}; \citealt{Gupta2019}; \citealt{Venturini2020}) requires planets with both precise radii and masses, a sample that remains limited.

How stellar multiplicity shapes the population of small planets is a question that has only recently started being studied. Using \textit{Kepler} planet hosts with directly imaged stellar companions, \citet{Sullivan2023, Sullivan2024} found that the radius distribution of small S-type planets (planets orbiting one star of the binary) depends on the binary separation: wide binaries ($a_{\rm bin}$>300 au) and single stars host indistinguishable small-planet radius distributions, while close binaries ($a_{\rm bin}$<100 au) show a suppressed population of sub-Neptunes and no detected radius valley. This suppression was recently confirmed using a bias-corrected occurrence sample by \citet{Sullivan2026A}, who found $\sim$50\% fewer planets per system in close binaries than around single stars. However, these results rely on planets that lack precise radii and on the assumption that they are orbiting the primary star, since the faintness of the companions makes the true host difficult to identify. Enlarging the sample of well-characterized S-type small planets is therefore essential to understand whether the radius valley remains, shifts, or disappears in such systems, and whether its location is correlated with the binary separation.

Stellar companions are not the only external components that shape the architecture and composition of inner small-planet systems. Massive planetary companions on wide orbits can also play an important role. Whether outer giant planets favor or suppress the occurrence of inner super-Earths and sub-Neptunes is still unclear. Dynamical studies predict that a massive giant can scatter away close-in low-mass planets, so systems with giant companions should lack inner small planets (\citealt{Juric2008}; \citealt{Bitsch2020}). However, observations find a positive correlation between inner small planets and outer cold Jupiters (\citealt{ZhuWu2018}; \citealt{Bryan2019}; \citealt{Bryan2024}; \citealt{Bonomo2025}), although recent radial velocity (RV) searches seem to contradict this (\citealt{Delisle2025A}). Outer giants can also affect the composition of the inner planets, depending on when the giant forms and on whether the solids are mostly pebbles or planetesimals. Since pebbles drift inward from the outer disk, a giant traps them in the pressure bump beyond its orbit and the inner region is left poor in solids (e.g., \citealt{Lambrechts2019}). Planetesimals are accreted more locally, so a dry inner planet can still form inside the giant's orbit \citep{Schlecker2021}. If instead the inner planet forms beyond the ice line and migrates in before the giant, it can remain water-rich \citep{Venturini2020}. Increasing the sample of well-characterized inner sub-Neptunes with massive, long-period companions is therefore crucial to test these scenarios. When the host is also part of a stellar binary, the stellar and planetary companions both affect the inner planet, making these systems benchmarks for planet formation.

In this paper, we present the detection and characterization of the TOI-1699A S-type multiplanet system, in which both planets orbit the G5\,V primary of a wide stellar binary. We confirm and characterize the transiting sub-Neptune TOI-1699A b, initially detected by the Transiting Exoplanet Survey Satellite (TESS), using photometry from the CHaracterising ExOPlanet Satellite (CHEOPS) and ground-based photometry together with RVs from the High Accuracy Radial velocity Planet Searcher for the Northern hemisphere (HARPS-N). The RV data also reveal a massive companion, TOI-1699A c, on a wider orbit. In Sect.~\ref{sec:Observations} we present the photometric, spectroscopic, and imaging observations, while Sect.~\ref{sec:stellarcharacterization} describes the stellar properties of the binary. The system validation and global analysis are detailed in Sect.~\ref{sec:system}. Finally, the results are discussed in Sect.~\ref{sec:Discussion}, and our conclusions are summarized in Sect.~\ref{sec:Conclusions}.

\section{Observations}\label{sec:Observations}
\subsection{Multiplicity of the TOI-1699 system from Gaia}\label{sec:Gaia}
TOI-1699 was identified as a comoving binary system by \citet{Mugrauer2022} and \citet{Behmard2022} using \textit{Gaia} EDR3 (\citealt{GaiaEDR3}) astrometry and photometry. The system is located at a distance of 206.9 $\pm$ 1.3~pc and consists of the primary star TOI-1699A (\textit{Gaia} DR3 930290225551897984; TYC 3415-00428-1) and a faint companion, TOI-1699B (\textit{Gaia} DR3 930290229846761216), separated by 1.68~arcsec, corresponding to a projected separation of approximately 348~au. The secondary star has common proper motion and consistent parallax relative to the primary star, confirming that the two sources are gravitationally bound. TOI-1699B is fainter than the primary by $\Delta G$ = 5.25\,mag. Using the absolute \textit{Gaia} $G$-band magnitude and interpolated $G$-mass and $G$-$T_{\rm eff}$ relations from \citet{Pecaut2013}, \citet{Mugrauer2022} estimated TOI-1699B to have an effective temperature of $3368^{+18}_{-23}$~K, consistent with an M3\,V star. We derive and refine the stellar parameters for the primary and secondary components in Sect.~\ref{sec:stellarcharacterization}.

\subsection{TESS discovery photometry}\label{sec:TESS}
TOI-1699 (TIC 22705558) was initially observed by TESS (\citealt{Ricker2015}) in Sector 20 between 2019 December 24 and 2020 January 20 at 30-min cadence. After detection by the Quick Look Pipeline at Massachusetts Institute of Technology \citep{Huang2020a,Huang2020b}, TOI-1699A b was alerted as a Level 1 exoplanet candidate on 2020 February 19 by the TESS Science Office as part of the TESS Objects of Interest (TOI) catalog \citep{Guerrero2021}. The target was reobserved in Sector 47 from 2021 December 30 to 2022 January 28 and in Sector 60 from 2022 December 23 to 2023 January 18 at 2-min cadence. As of September 2026, no future TESS observations are scheduled. The light curves from Sectors 47 and 60 were produced by the Science Processing Operations Center (SPOC; \citealt{Jenkins2016}) at the NASA Ames Research Center from the target pixel files (TPFs), while the Sector~20 light curve was extracted from the full-frame images by the TESS-SPOC pipeline \citep{Caldwell2020}. The SPOC pipeline provides both the simple aperture photometry (SAP; \citealt{Twicken2010}; \citealt{Morris2020}) and presearch data conditioning simple aperture photometry \citep[PDCSAP;][]{Stumpe2012,Stumpe2014,Smith2012} products, where SAP corresponds to the raw flux, while PDCSAP is corrected for instrumental systematics and flux dilution. For the host assignment prior to the RV follow-up we used the SAP flux (Sect.~\ref{sec:host}), while for our final joint analysis we used the PDCSAP flux (Sect.~\ref{sec:join}).

Given the large TESS pixel size (21~arcsec~pixel$^{-1}$) and the binary nature of the system, we examined the TESS TPF of TOI-1699 using \texttt{tpfplotter} \citep{TPFplotter}. The only two \textit{Gaia} DR3 sources within the adopted TESS aperture are TOI-1699A and TOI-1699B, while
the contribution from the other nearby sources is minimal. Despite its proximity, the flux contribution of TOI-1699B is minor but nonnegligible ($\Delta T \approx 5.27$\,mag). To ensure accurate planet parameters, our final analysis accounts for this contamination by using the dilution-corrected PDCSAP photometry (Sect.~\ref{sec:join}). The TESS light curves are shown in Fig.~\ref{fig:TESS_full}.

\subsection{CHEOPS targeted photometric follow-up}\label{sec:CHEOPS}
\begin{figure*}[t!]
  \centering
  \includegraphics[width=1.\textwidth]{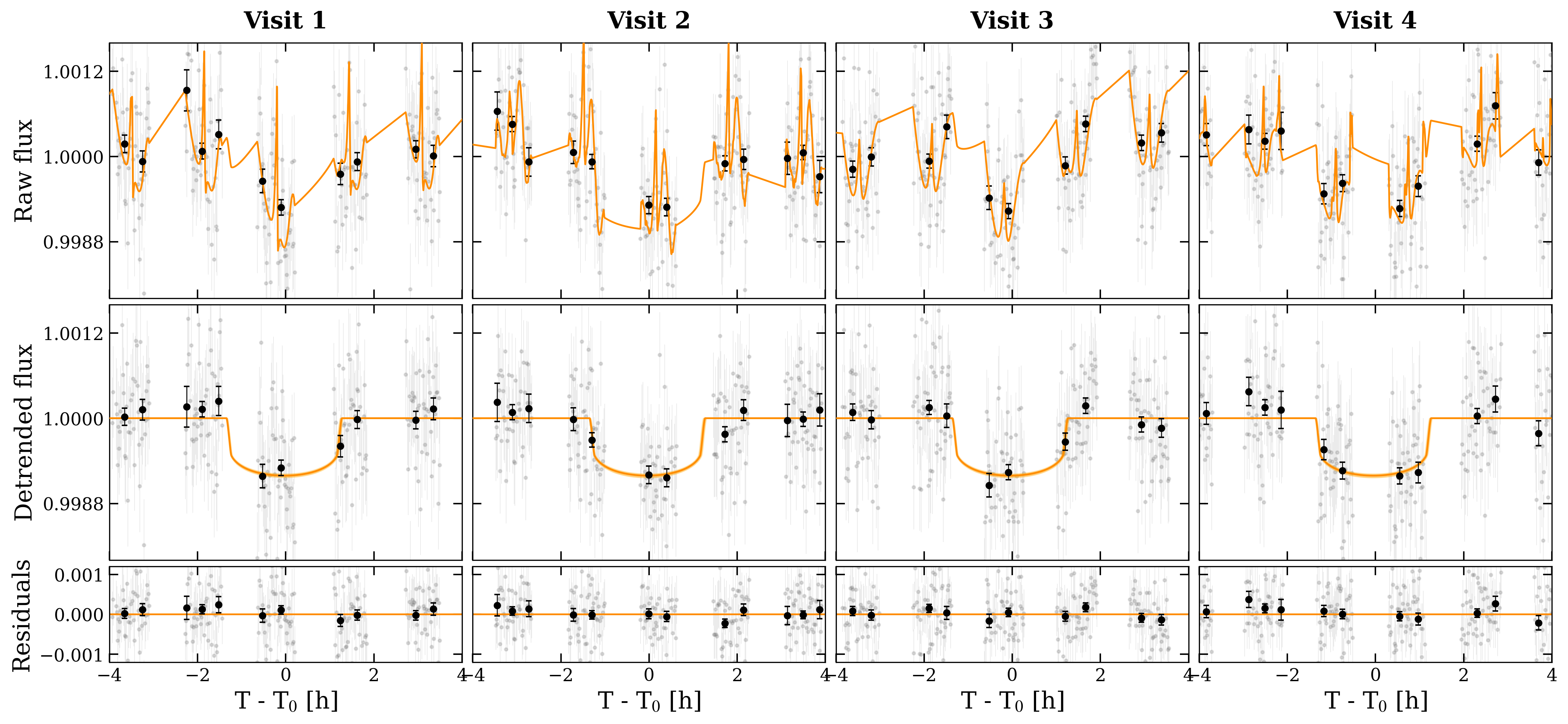}
  \caption{CHEOPS light curves of TOI-1699A b from four visits (Table \ref{tab:cheopsobservations}). The top panels show the raw flux and best-fit global model (orange line), middle panels show the detrended light curves with the best-fit transit model (orange line) and its $\pm$ 1$\sigma$ region (shaded orange), and bottom panels show the residuals. Gray points correspond to the unbinned data, while black points show the binned data.}
  \label{fig:cheopslcs}
\end{figure*}

We performed targeted CHEOPS (\citealt{Benz2021}; \citealt{Fortier2024}) photometric observations for the transiting inner planet TOI-1699A b as part of the S-Valley CHEOPS Guaranteed Time Observation (GTO) program (PI: Venturini; CH$\_$PR140085; Venturini et al., in prep.). The purpose of the observations was to improve the transit depth precision, to refine the stellar density for the host-star identification (Sect.~\ref{sec:host}), and to foster RV follow-up efforts (Sect.~\ref{sec:harps}), given that the target will not be reobserved by TESS. In total, we obtained four visits between 2024 January 10 and 2025 February 9 with an exposure time of 60 s and efficiencies from 57 to 58\%. The CHEOPS photometry was extracted with {\tt PIPE} (PSF Imagette Photometric Extraction; \citealt{Brandeker2024}), which performs point-spread-function (PSF) fitting rather than the aperture photometry of the default Data Reduction Pipeline ({\tt DRP}; \citealt{Hoyer2020}). This makes it better suited than {\tt DRP} for such closely separated binaries, as it reduces contamination and associated roll-angle systematics and thus results in lower scattered light. At a separation of 1.68~arcsec, TOI-1699B is fully blended with the primary. {\tt PIPE} corrects for it using the companion's \textit{Gaia} flux ratio, and because the residual contamination scales as the square of the flux ratio, the effect on the transit depth is negligible. The observations are listed in Table~\ref{tab:cheopsobservations}, the detrending method is described in Sect.~\ref{sec:join}, while the raw and detrended light curves are shown in Fig.~\ref{fig:cheopslcs}.

\begin{table*}[t!] 
\small
\caption{CHEOPS photometric observations of TOI-1699A b.}
\centering
\renewcommand{\arraystretch}{1.1}
\setlength{\tabcolsep}{6pt}

\begin{tabular}{cccccclc}
\hline\hline
Visit & Start date & Duration & Efficiency & Frames & File key & Detrending & Planet \\
 & [UTC] & [h] & [$\%$] & [No.] & & & \\
\hline
1 & 2024 January 10 23:44 & 13.71 & 57 & 478 & CH\_PR140085\_TG001001\_V0300 & $t,\, \mathrm{spline}_3(\phi)$ & b \\
2 & 2024 March 10 03:01 & 14.79 & 58 & 512 & CH\_PR140085\_TG000101\_V0300 & $\mathrm{bg},\, \mathrm{cont}^2,\, \mathrm{spline}_3(\phi)$ & b \\
3 & 2025 January 13 07:18 & 13.24 & 57 & 461 & CH\_PR140085\_TG008101\_V0300 & $t,\, \mathrm{spline}_3(\phi)$ & b \\
4 & 2025 February 9 14:58 & 14.07 & 58 & 515 & CH\_PR140085\_TG008001\_V0300 & $\mathrm{bg},\, \mathrm{spline}_3(\phi)$ & b \\
\hline
\end{tabular}

\begin{tablenotes}
\item \textbf{Notes:}
Detrending terms are given as functions of the corresponding parameter and the polynomial orders are indicated by powers. Here, $t$, $\mathrm{bg}$, $\mathrm{cont}$, and $\phi$ denote time, background, contamination, and roll angle, respectively. 
$\mathrm{spline}_3(\phi)$ denotes a cubic spline as a function of the roll angle $\phi$.
\end{tablenotes}

\label{tab:cheopsobservations}
\end{table*}

\subsection{Ground-based photometric follow-up}\label{sec:ground}
\begin{figure}[t!]
  \centering
  \includegraphics[width=0.5\textwidth]{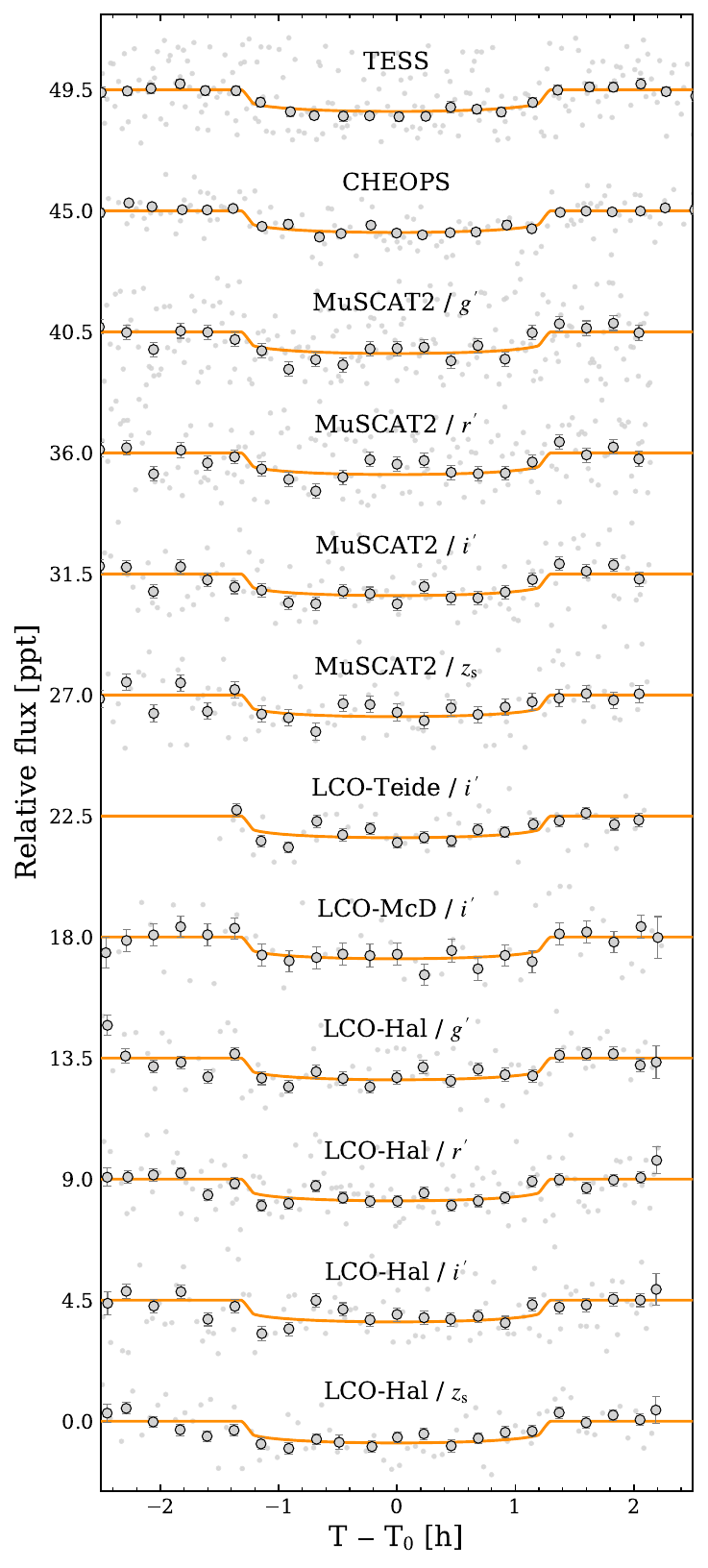}
\caption{Phase-folded light curves of TOI-1699A b from TESS, CHEOPS, and 
the ground-based instruments. Light gray 
points show the individual measurements, darker points the binned data, 
and orange lines the best-fit transit model from the joint analysis.}
\label{fig:stackedlcs}
\end{figure}

We obtained ground-based observations to confirm that the transit signal of TOI-1699A b originates from the target star and to check that the transit depth is consistent across different bands. We used the {\tt TESS Transit Finder}, which is a customized version of the {\tt Tapir} software package\footnote{\url{https://github.com/elnjensen/Tapir}}, to schedule our transit observations. All follow-up photometric apertures described below include flux from TOI-1699B. All observations confirmed that the event is on-target and are included in the joint analysis described in Sect. \ref{sec:join}. The light curves are presented in Fig. \ref{fig:stackedlcs}.

We observed a full transit window of TOI-1699A b in Sloan $i'$ on 2023 November 13 from the Las Cumbres Observatory Global Telescope \citep[LCOGT;][]{Brown:2013} 1.0\,m network node at Observatorio del Teide on the island of Tenerife (Teide). Another full transit window was observed in Sloan $i'$ on 2023 December 1 from the 1.0\,m network node at McDonald Observatory near Fort Davis, Texas, United States (McD). LCOGT 1\,m network telescopes are equipped with a $4096~\times~4096$ SINISTRO camera with an image scale of 0.389~arcsec~pixel$^{-1}$, resulting in a 26~arcmin $\times$ 26~arcmin field of view. The Teide and McD images were calibrated by the standard LCOGT {\tt BANZAI} pipeline \citep{McCully:2018}, and differential photometric data were extracted using {\tt AstroImageJ} \citep{Collins:2017}. We used circular photometric apertures with radii of 7~arcsec and 8.6~arcsec for Teide and McD, respectively, which included the flux from TOI-1699B.

On 2024 November 28, the target TOI-1699A was observed using the MuSCAT2 multiband imager \citep{Narita2019}, which is mounted on the 1.5~m Telescopio Carlos S\'{a}nchez at Teide Observatory, Spain. MuSCAT2 is equipped with four CCDs capable of obtaining simultaneous images in the $g'$, $r'$, $i'$, and $z_s$ bands with negligible readout time. Each CCD consists of $1024 \times 1024$ pixels, providing a field of view of 7.4~arcmin $\times$ 7.4~arcmin. The observations were conducted with the telescope slightly defocused to prevent target saturation. The exposure times were set to 10~s for $g'$, 15~s for $r'$, and 25~s for both $i'$ and $z_s$. The raw data were reduced using the MuSCAT2 pipeline \citep{Parviainen2019}, which performs standard dark- and flat-field calibrations, aperture photometry, and a joint transit and instrumental systematics model fit. 

We also observed a full transit window of TOI-1699A b on 2024 December 26 simultaneously in Sloan $g'$, $r'$, $i'$, and Pan-STARRS $z_s$ from the LCOGT 2\,m Faulkes Telescope North at Haleakala Observatory on Maui, Hawai'i. The 2\,m telescope is equipped with the MuSCAT3 multiband imager \citep{Narita:2020}. All images were calibrated by the standard LCOGT {\tt BANZAI} pipeline \citep{McCully:2018} and differential photometric data were extracted using {\tt AstroImageJ} \citep{Collins:2017}. Circular apertures with radii of 5.3~arcsec were used to extract the differential photometry. All apertures included flux from TOI-1699B, as the two stars could not be separated from one another in the photometric apertures. The MuSCAT3 observations across Sloan $g'$, Sloan $r'$, Sloan $i'$, and $z_s$ bands show no indication of wavelength-dependent depth changes in the transits of TOI-1699A b.

\subsection{High-resolution spectroscopy with HARPS-N}\label{sec:harps}
We performed high-resolution spectroscopic follow-up with HARPS-N (\citealt{Cosentino2012}) within the CAT25B$\_$53 program (PI: Psaridi). HARPS-N is mounted on the 3.58 m Telescopio Nazionale Galileo telescope at the Observatorio del Roque de los Muchachos, Spain, and is fed by a science fiber with an on-sky diameter of 1~arcsec. The spectra were collected between 2025 October 8 and 2026 March 28 with an exposure time of 1800~s. A total of 28 spectra were obtained, two of which were discarded due to poor seeing conditions, resulting in 26 spectra used in the analysis. The retained spectra were collected with a median seeing of $\sim$1.2~arcsec. We extracted the RVs and activity indicators (FWHM, contrast, and bisector) with version 3.2.0 of the HARPS-N Data Reduction Software (\citealt{Dumusque2021}) using a G2 binary mask (\citealt{Pepe2002}). The spectra have signal-to-noise ratios of S/N = 28–53 at 550~nm, and the RVs exhibit an rms scatter of 132~\ms with a median uncertainty of 2.4~\ms. The RVs and associated activity indicators are presented in Table~\ref{tab:rvs}.

\subsection{High-resolution imaging}\label{sec:imaging}
\begin{figure}[t!]
  \centering
  \includegraphics[width=.50\textwidth]{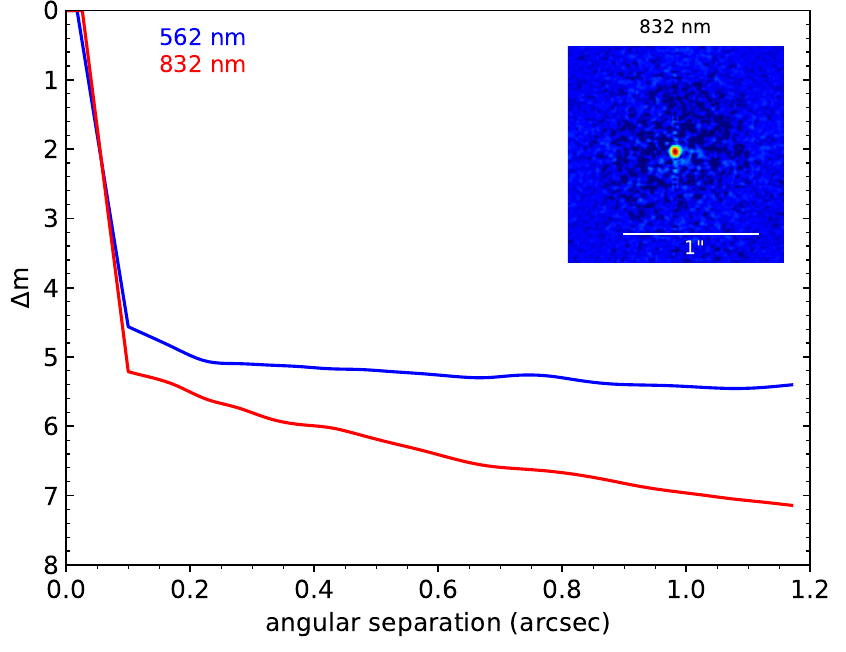}
\caption{5$\sigma$ contrast curves for TOI-1699A from the 'Alopeke speckle imager at Gemini North. The inset shows the reconstructed 832\,nm image, where the horizontal bar indicates 1~arcsec.} 
\label{fig:gemini}
\end{figure}

Apart from the known stellar companion TOI-1699B at 1.68~arcsec, high-resolution imaging is needed to rule out closer companions that could
dilute the transit signal. Therefore, we obtained speckle imaging of TOI-1699A with the 'Alopeke instrument \citep{Scott2021} on the Gemini North telescope on 2021 February 2. 'Alopeke provides simultaneous speckle imaging in two bands; for the TOI-1699A observations, the narrow-band 562 nm and 832 nm filters were chosen, and the star was observed in six sets of 1000 $\times$ 0.06 s images. The data were processed with the standard speckle reduction pipeline \citep{Howell2011}, whose output data products include 5$\sigma$ magnitude contrast curves and reconstructed images, showing a $\sim$1.2~arcsec field of view around the target and potentially resolving companions down to the diffraction limit of the Gemini telescope. We detect no companions within 0.1--1.2~arcsec down to contrasts of 5--7 magnitudes at 832 nm (Fig. \ref{fig:gemini}). TOI-1699 was also observed with the AstraLux lucky imager at the Calar Alto 2.2\,m telescope on 2020 February 25 in the SDSS $z$ band (sensitivity of 4.7~mag at 1~arcsec) with no detected companion \citep{LilloBox2024}.

\section{Properties of the host star}\label{sec:stellarcharacterization}
\subsection{Atmospheric parameters and abundances}
We derived the stellar spectroscopic parameters ($T_{\rm eff}$, $\log{g}$, microturbulence, [Fe/H]) using the {\tt ARES}+{\tt MOOG} methodology \citep{2021A&A...656A..53S, 2014dapb.book..297S, 2013A&A...556A.150S}. We used the {\tt ARES}\footnote{The latest version, {\tt ARES} v2, can be downloaded at \url{https://github.com/sousasag/ARES}} code \citep{2007A&A...469..783S,2015A&A...577A..67S} to analyze the co-added HARPS-N spectrum of TOI-1699A (S/N $\approx$ 220 per pixel at 550 nm) to measure the equivalent widths (EWs) for a list of lines presented by \cite{2008A&A...487..373S}. The best set of spectroscopic parameters was found by using a minimization process to determine the ionization and excitation equilibrium. The latter process relies on a grid of Kurucz model atmospheres \citep{Kurucz1993} and the most recent version of the radiative transfer code, {\tt MOOG} \citep{1973PhDT.......180S}. We derived a more accurate surface gravity using \textit{Gaia} DR3 data, following the same procedure as \cite{2021A&A...656A..53S}. This value is consistent with the spectroscopic estimate.

We also determined the abundances of Mg and Si using the atmospheric parameters and the classical curve-of-growth method \citep[][]{Adibekyan-15, Adibekyan-16}. The EWs of the spectral lines were measured with {\tt ARES}, and a careful visual inspection of the line fitting did not reveal any apparent problems. Using the same grid of model atmospheres and radiative transfer code described above, we converted the EWs into elemental abundances under the assumption of local thermodynamic equilibrium (Table~\ref{tab:Stellar parameters}). The star exhibits an abundance pattern typical of thin-disk stars in the solar neighborhood \citep[][]{Adibekyan-12}.

\subsection{Infrared flux method and isochrones}
\label{sec:irfm}
From the atmospheric parameters ($T_{\mathrm{eff}}$, $\log{g}$, [Fe/H]), we derived the stellar radius, mass, and age of TOI-1699A using a custom-built code that combines the infrared flux method \citep{1977MNRAS.180..177B, 2017AJ....153..136S} and isochrone fitting \citep[{\tt MCMCI} code,][]{2020A&A...635A...6B}. Our approach is fully described by \cite{Bonfanti2026} and \cite{Mantovan2026}. We obtained $R_{\star,\rm A}$ = 1.036 $\pm$ 0.009~$R_\odot$, $M_{\star,\rm A} = 0.989_{-0.054}^{+0.052}~M_\odot$, and an age of $5.4_{-1.4}^{+2.1}$~Gyr. The stellar parameters are reported in Table \ref{tab:Stellar parameters}. We repeated the analysis for the secondary star TOI-1699B, this time starting only from \textit{Gaia} astrometry and photometry ($G$, $G_{\mathrm{BP}}$, $G_{\mathrm{RP}}$), as well as the stellar temperature $T_{\rm eff}$ estimated by \cite{Mugrauer2022}. We found $M_{\star,\rm B} = 0.401_{-0.033}^{+0.026}~M_\odot$, $R_{\star,\rm B} = 0.407_{-0.032}^{+0.037}~R_\odot$, and $\rho_{\star,\rm B} = 8.2_{-2.5}^{+2.0}$~g~cm$^{-3}$.
\begin{table}[t!]
\small
\caption{Stellar parameters for TOI-1699A.}
\centering
\renewcommand{\arraystretch}{1.1}
\begin{tabular}{lcc}
\hline\hline
  & TOI-1699A  & Source \\
    \hline
    \textit{Identifiers}& &\\
    TIC ID  & 22705558 & TICv8\\
    2MASS ID & J08204531+4612134 &  2MASS\\
    \textit{Gaia} ID & 930290225551897984 & \textit{Gaia} DR3\\
    TYC & 3415-00428-1 & Tycho-2\\
    \hline
    \textit{Astrometric parameters} & &\\
    $\alpha$ (J2016) & 08:20:45.30 & \textit{Gaia} DR3\\
    $\delta$ (J2016) & +46:12:13.0 & \textit{Gaia} DR3\\
    $\pi$ (mas) & 4.834 $\pm$ 0.032 & \textit{Gaia} DR3\\
    $d$ (pc) & 206.9 $\pm$ 1.3 & \textit{Gaia} DR3\\
    $\mu_\alpha \cos\delta$ (mas yr$^{-1}$) & $-$5.182 $\pm$ 0.035 & \textit{Gaia} DR3\\
    $\mu_\delta$ (mas yr$^{-1}$) & $-$25.461 $\pm$ 0.027 & \textit{Gaia} DR3\\
    RUWE & 1.46 & \textit{Gaia} DR3\\
    \hline
    \textit{Photometric parameters} & &\\
    $B_T$ (mag) & 12.21 $\pm$ 0.20  & Tycho-2\\
    $V_T$ (mag) & 11.48 $\pm$ 0.13 & Tycho-2\\
    $G$ (mag) & 11.3265 $\pm$ 0.0028 & \textit{Gaia} DR3 \\
    $T$ (mag) & 10.8756 $\pm$ 0.0061 & TICv8\\
    $J$ (mag) & 10.251 $\pm$ 0.022 & 2MASS\\
    $H$ (mag) & 9.912 $\pm$ 0.016 & 2MASS\\
    $K$ (mag) & 9.852 $\pm$ 0.018& 2MASS\\
    $W1$ (mag) & 9.806 $\pm$ 0.023 & WISE\\
    $W2$ (mag) & 9.841 $\pm$ 0.020& WISE\\
    $W3$ (mag) & 9.726 $\pm$ 0.043& WISE\\
    $W4$ (mag) & 8.51 $\pm$ 0.29 & WISE\\
    \hline
    \textit{Bulk parameters} & &\\
    Spectral type & G5\,V & Sect. \ref{sec:stellarcharacterization} \\
    \teff\,(K) & 5649 $\pm$ 62  &Sect. \ref{sec:stellarcharacterization} \\
    $R_\star$  (\rsol)  & 1.036 $\pm$ 0.009  & Sect. \ref{sec:stellarcharacterization}\\
    $M_\star$  (\msol)   & 0.989$^{+0.052} _{-0.054}$ &Sect. \ref{sec:stellarcharacterization} \\
    $\rho_\star$  (\denssol)   & 0.879$^{+0.053} _{-0.054}$ &Sect. \ref{sec:stellarcharacterization} \\
    \feh (dex)  & 0.10 $\pm$ 0.04 &Sect. \ref{sec:stellarcharacterization}\\
    \sih (dex)  & 0.14 $\pm$ 0.04 &Sect. \ref{sec:stellarcharacterization}\\
    \mgh (dex)  & 0.10 $\pm$ 0.05 &Sect. \ref{sec:stellarcharacterization}\\
    $\log g_\star$ (cgs)  & 4.45 $\pm$ 0.03 &Sect. \ref{sec:stellarcharacterization} \\
    Age	(Gyr)  & 5.4$^{+2.1} _{-1.4}$ &Sect. \ref{sec:stellarcharacterization}\\
    $E(B-V)$ (mag)  & 0.017 $\pm$ 0.006&Sect. \ref{sec:stellarcharacterization}\\
   \hline
\end{tabular}
\\
\begin{tablenotes}
\item \textbf{Sources:} 1. TICv8 \citep{Stassun2019}; 2. 2MASS \citep{Skrutskie2006}; 3. \textit{Gaia} DR3 \citep{GAIADR3}; 4. Tycho-2 \citep{Tycho}; 5. WISE \citep{Wright2010}.
\end{tablenotes}
\label{tab:Stellar parameters}
\end{table}

\section{System discovery and characterization}\label{sec:system}
\subsection{Stellar host assignment}\label{sec:host}
Prior to the RV follow-up, we performed a stellar host-identification analysis using the TESS SAP and the CHEOPS {\tt DRP} photometry following the method described by Venturini et al. (in prep.). We considered two scenarios, assuming that the transiting planet orbits either the primary or the secondary star, and applied the corresponding dilution corrections to the photometry in each case. The dilution factors were computed from all nearby stars located within the photometric apertures. The primary-host scenario requires only small dilution corrections ($D_\mathrm{TESS} = 0.008$, $D_\mathrm{CHEOPS} = 0.005$), while the secondary-host scenario requires much larger factors ($D_\mathrm{TESS} = 128$, $D_\mathrm{CHEOPS} = 201$). In both cases, we fitted for the stellar density using wide uniform priors ranging from 0.1 to 20~g~cm$^{-3}$. The stellar densities derived from the transit fit were then compared with the stellar densities inferred from stellar models (Sect.~\ref{sec:stellarcharacterization}). Our analysis favored the primary star TOI-1699A as the planet-hosting star (Table~\ref{tab:densities}), a result that was later confirmed with our HARPS-N RV follow-up.

\begin{table}[t!]
\caption{Transit-derived and model-derived stellar densities under the two host-star hypotheses.}
\centering
\small
\renewcommand{\arraystretch}{1.5}
\begin{tabular*}{\columnwidth}{@{\extracolsep{\fill}}lcc}
\hline
\hline
Host 
& $\rho_{\star,\mathrm{transit}}$ (g~cm$^{-3}$)  
& $\rho_{\star,\mathrm{model}}$ (g~cm$^{-3}$) \\
\hline
Primary   
& 1.24$^{+0.29}_{-0.31}$
& $1.24^{+0.07}_{-0.08}$ \\
Secondary 
& $2.38^{+0.51}_{-0.54}$ 
& 8.2$^{+2.0}_{-2.5}$\\
\hline
\end{tabular*}
\label{tab:densities}
\end{table}

\subsection{Detection of the outer planet}\label{sec:outer}
\begin{figure}[t!]
  \centering
  \includegraphics[width=.50\textwidth]{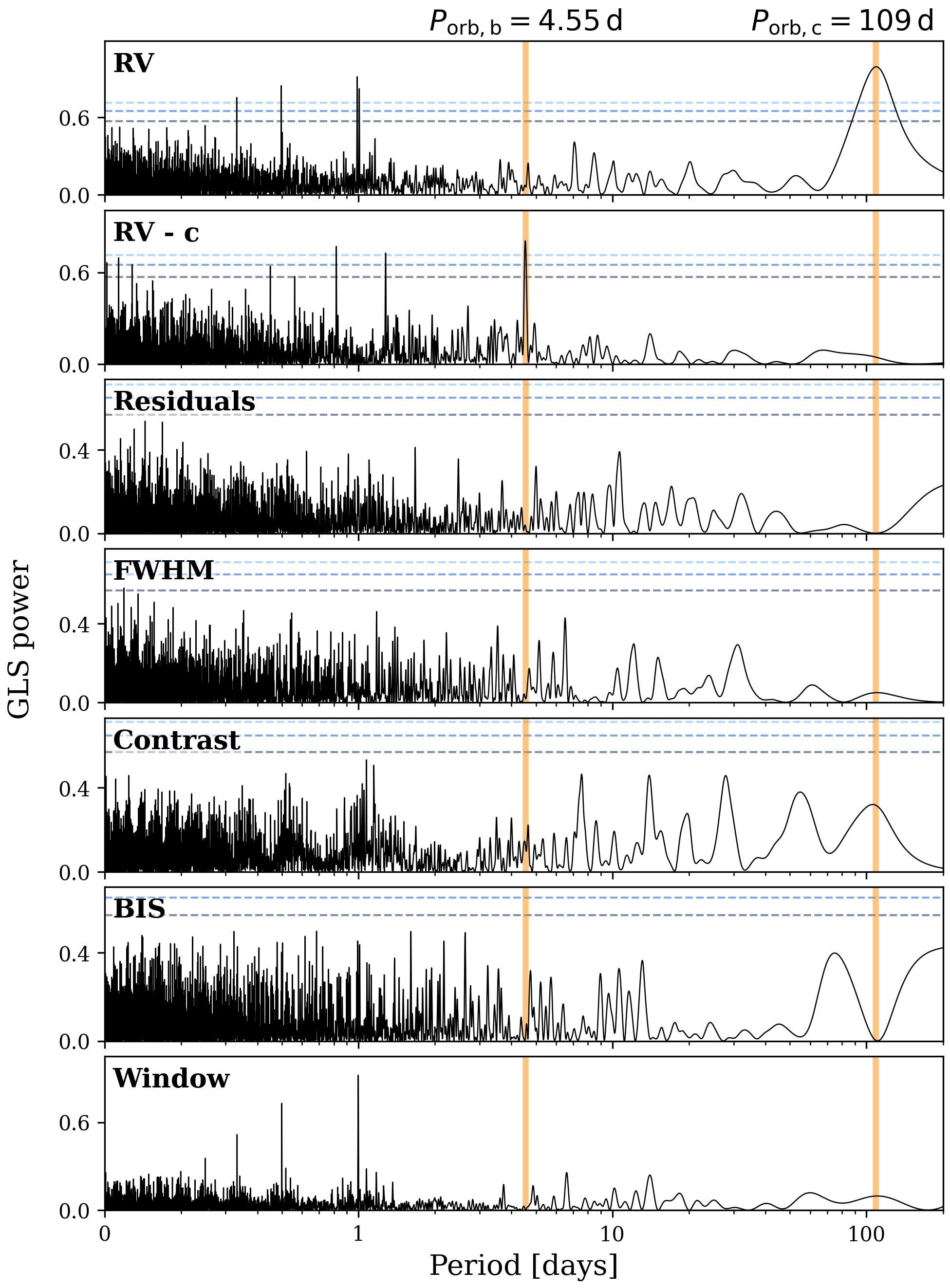}
\caption{{\tt GLS} periodogram of the HARPS-N RVs of TOI-1699A, corresponding 
activity indicators, and window function. Horizontal dashed lines indicate the FAP levels of 10\%, 1\%, and 0.1\%, shown from darker to lighter blue. The vertical orange lines mark the orbital periods of TOI-1699A b and TOI-1699A c.}
\label{fig:rvgls}
\end{figure}

TOI-1699A c was detected in the HARPS-N data using a generalized Lomb--Scargle periodogram ({\tt GLS}; \citealt{Zechmeister2009}). As shown in Fig. \ref{fig:rvgls}, the {\tt GLS} reveals a prominent and isolated peak at $P \approx 109$ days, which dominates the power spectrum. This signal is highly significant, with a false alarm probability (FAP) of $\sim7 \times 10^{-22}$, well below the 0.1$\%$ threshold. We verified that the signal is not related to stellar activity by computing {\tt GLS} periodograms of common activity indicators (FWHM, contrast, and bisector), and the window function, none of which show significant power at that frequency. No significant power connected to stellar rotation is present in the activity indicators or the RV residuals, and the TESS SAP photometry shows no significant rotational modulation. Therefore, we did not include a stellar-activity term in our final model. Moreover, we checked whether the signal originates from the primary or the companion star. With a fiber aperture of 1~arcsec and a median seeing of $\sim$1.2~arcsec (Sect.~\ref{sec:harps}), the flux from TOI-1699B is strongly suppressed. TOI-1699B is located at a separation of 1.68~arcsec, beyond the fiber radius, and is 5.25\,mag fainter than the primary one, corresponding to a flux ratio of only $\sim0.8\%$. Such a diluted blend could only reproduce the observed semi-amplitude of $K_{\rm c} = 238$~m\,s$^{-1}$ if TOI-1699B varied by tens of km\,s$^{-1}$, which would be evident in the bisector. This indicates that TOI-1699A c orbits TOI-1699A as well. After fitting a Keplerian model to the outer companion and subtracting its signal from the RVs, a second peak at $P \sim4.55$ days becomes apparent in the residuals, consistent with the orbital period of the transiting inner planet identified in the TESS photometry. This signal is also significant, with a FAP of $\sim1 \times 10^{-5}$. No additional significant signals are detected in the residuals after subtracting both planetary signals. The transiting configuration and estimated properties of TOI-1699A c are discussed in Sect.~\ref{sec:TOI-1699A c}.

\subsection{Joint photometric and RV analysis}\label{sec:join}
We jointly modeled the TESS, CHEOPS, and ground-based photometric data together with the HARPS-N RVs using the COde for exoplaNet ANalysis (\CONAN; \citealt{Lendl2017}; \citealt{Akinsanmi2025}). We used the PDCSAP TESS light curves, which are already corrected for contamination, and the {\tt PIPE} CHEOPS photometry, for which contamination from background sources is taken into account during the PSF extraction. For the ground-based multiband photometry, we evaluated the expected flux dilution from the secondary star, TOI-1699B. While the companion is 5.25\,mag fainter in the \textit{Gaia} $G$ band, its M3\,V spectral type yields a significantly redder spectral energy distribution. Using the empirical dwarf sequence from \citet{Pecaut2013} transformed to SDSS magnitudes via \citet{Jester2005}, we estimate the flux contamination to be negligible in the optical ($<0.5\%$ in the $g'$ and $r'$ bands), rising to $\sim1.1\%$ in the $i'$ band and $\sim2.2\%$ in the $z_s$ band. Because the photometric precision of our ground-based light curves is dominated by scatter and environmental systematics exceeding these contamination levels, we did not apply a dilution correction to the ground-based data.

In our analysis we fitted the following jump parameters: stellar density ($\rho_*$), orbital period ($P$), time of conjunction ($T_{\mathrm{c}}$), ratio of planetary to stellar radii ($R_{\mathrm{P}}/R_{\mathrm{*}}$), impact parameter ($b$), limb-darkening coefficients, RV amplitude ($K$), eccentricity ($e$), and argument of periastron ($\omega$). The stellar density was fitted with a Gaussian prior based on the value derived from the stellar modeling and its uncertainty (Sect.~\ref{sec:stellarcharacterization}). The quadratic limb-darkening coefficients (\citealt{Kipping2013}) were also fitted with Gaussian priors derived from the stellar parameters with \ldtk \citep{Parviainen2015}. As shown in Table \ref{tab:posteriors}, for the rest of the parameters, we adopted wide Gaussian or uniform priors centered on the signals identified by the TESS SPOC pipeline and the {\tt GLS} periodogram of the HARPS-N RVs.

The correlated variability in the TESS photometry was corrected using Gaussian processes (GPs) with a Matérn-3/2 kernel (Fig. \ref{fig:TESS_full}). The CHEOPS and ground-based photometry were corrected through polynomial decorrelation against selected instrumental and environmental parameters (Fig. \ref{fig:cheopslcs}). The parameters and polynomial orders were selected using a forward-selection model, in which they were included only if they improved the fit according to the Bayesian Information Criterion (BIC; $\Delta\mathrm{BIC} \leq -5$). Additional white-noise terms were added in quadrature to the uncertainties of the photometric and RV time series.

To evaluate the significance of the eccentricity of TOI-1699A b, we first fitted the data allowing it to vary freely. For TOI-1699A b, with an eccentricity of $e = 0.055^{+0.066}_{-0.039}$, the Lucy--Sweeney criterion ($e > 2.45~\sigma_e$; \citealt{Lucy1971}) is not satisfied, indicating that the eccentricity is not significantly detected. In addition, a comparison of the Bayesian evidence values strongly favors a circular model, with $\ln Z_{\mathrm{circ}}$ = 197637.5 $\pm$ 0.6 and $\ln Z_{\mathrm{ecc}}$ = 197621.7 $\pm$ 0.6, corresponding to $\Delta \ln Z$ = 15.8 $\pm$ 0.8 in favor of the circular model ($\Delta \ln Z$ $>$ 5; \citealt{kass95}). We therefore adopt a circular orbit for the final fit and report the $3\sigma$ upper limit of $e < 0.27$. The priors we used as well as the median values and 68$\%$ confidence intervals of the posterior distributions are presented in Table \ref{tab:posteriors}. The best-fit RV model and the phase-folded signals of both planets are shown in Fig. \ref{fig:rvs}.

\begin{figure*}[t!]
  \centering
  \includegraphics[width=1.0\textwidth]{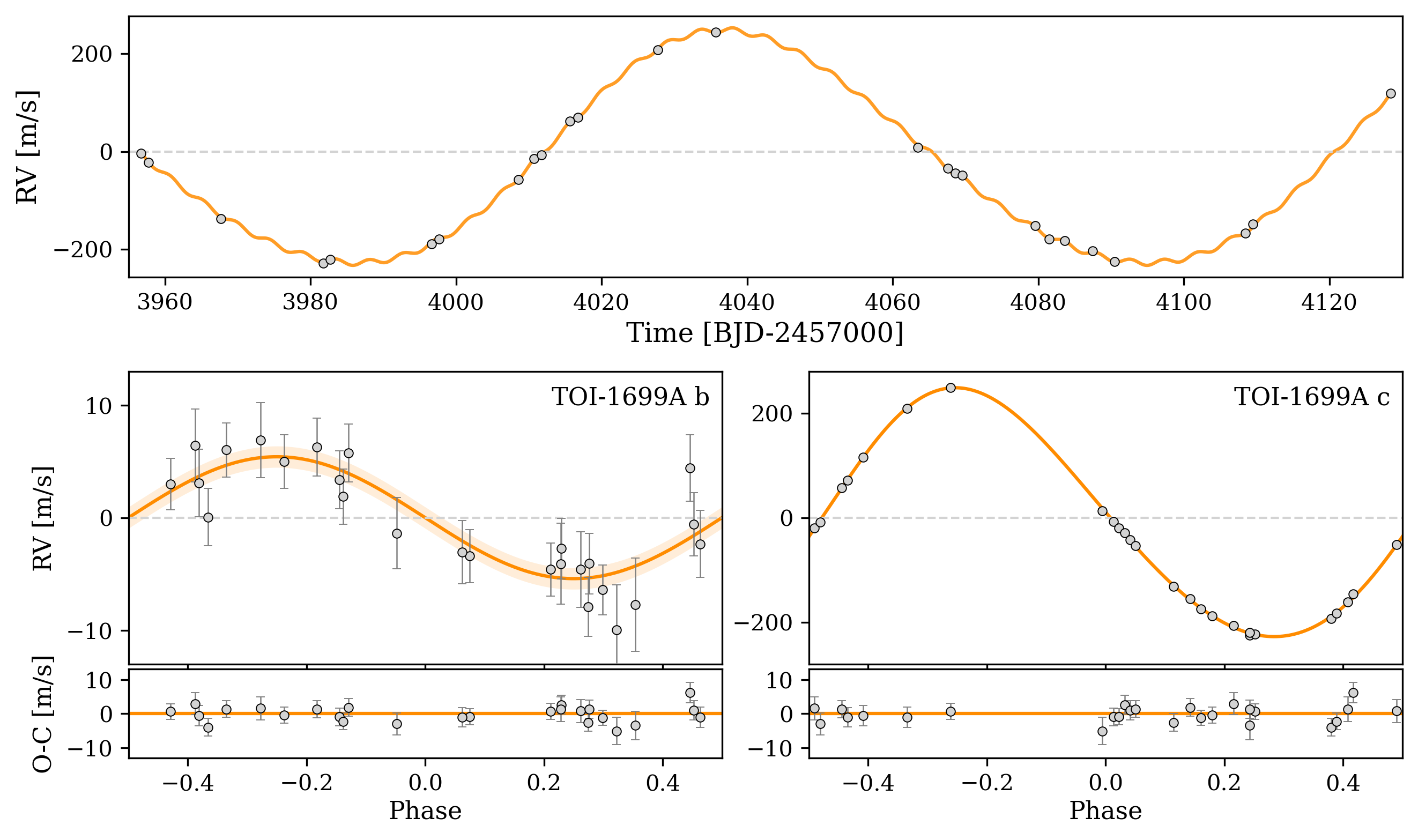}
\caption{HARPS-N RV observations and best-fit models for the TOI-1699A system. The top panel shows the full RV time series, while the lower panels show the phase-folded RVs of TOI-1699A b and TOI-1699A c with their residuals. The orange line indicates the best-fit model and the shaded orange region the $\pm$1$\sigma$ confidence interval.}
\label{fig:rvs}
\label{fig:rvs}
\end{figure*}

\begin{table}[t!]
\small
\caption{Jump and derived parameters for TOI-1699A b and TOI-1699A c.}
\centering
\renewcommand{\arraystretch}{1.5}
\setlength{\tabcolsep}{1.0pt}
\begin{center}
\begin{tabular*}{\columnwidth}{@{\extracolsep{\fill}}lcc}
\hline\hline
Parameter &TOI-1699A b & TOI-1699A c \\
\hline
Orbital period, $\it{P_{\mathrm{orb}}}$ (days) &$4.5470531^{+0.0000052}_{-0.0000052}$& $108.75^{+0.10}_{-0.10}$ \\
Time of conjunction, $T_\mathrm{c}$ (BTJD) & $2606.9075^{+0.0012}_{-0.0013}$ & $2867.8^{+1.1}_{-1.1}$ \\
Planet radius, $\it{R_{\mathrm{P}}}$ (\re) & $3.096^{+0.057}_{-0.056}$ & ... \\
Planet mass, $\it{M_{\mathrm{P}}}$ (\me) & $13.9^{+1.9}_{-1.8}$ & ... \\
Minimum mass, $\it{M_{\mathrm{P}}}\sin i$ (\mjup) & ... & $5.54^{+0.21}_{-0.21}$ \\
Planet density, $\rho_\mathrm{P}$ (g~cm$^{-3}$) & $2.58^{+0.40}_{-0.39}$ & ... \\
Orbital inclination, $i$ (deg) & $86.88^{+0.17}_{-0.17}$ & ... \\
Impact parameter, $\textit{b}$ & $0.603^{+0.024}_{-0.024}$ & ... \\
Semimajor axis, $\it{a}$ (au) & $0.0536^{+0.0010}_{-0.0010}$ & $0.4445^{+0.0084}_{-0.0084}$ \\
Transit duration, $T_{\mathrm{14}}$ (hours) & $2.618^{+0.028}_{-0.035}$ & ... \\
Scaled semimajor axis, $\it{a/\rstar}$ & $11.07^{+0.18}_{-0.18}$ & $91.9^{+1.5}_{-1.5}$ \\
Eccentricity, $e$ & $0$ ($<0.27$, 3$\sigma$) & $0.0737^{+0.0034}_{-0.0034}$ \\
Argument of periastron, $\omega_*$ (deg) & $90$ (fixed) & $307.1^{+2.8}_{-2.6}$ \\
Equilibrium temperature, $T_{\mathrm{eq}}$ (K) & $1200^{+18}_{-18}$ & $416.6^{+6.4}_{-6.3}$ \\
Atmospheric scale height, $H$ (km) & $303^{+47}_{-38}$ & ... \\
Escape velocity, $v_{\rm esc}$ (km\,s$^{-1}$) & $23.7^{+1.6}_{-1.6}$ & ... \\
Jeans escape parameter, $\Lambda$ & $28.3^{+3.9}_{-3.7}$ & ... \\
Insolation, $\it{S_{\mathrm{p}}}$ (\se) & $347^{+21}_{-20}$ & $5.03^{+0.30}_{-0.29}$ \\
\hline
\end{tabular*}
\begin{tablenotes}
\item
Notes: BTJD is the Barycentric TESS Julian Date (BJD$-$2457000). The equilibrium temperature $T_{\mathrm{eq}}$ (assuming a Bond albedo $A_{\mathrm{B}}=0$), the atmospheric scale height $H$ (for a mean molecular weight $\mu=2.3$; H/He-dominated atmosphere), the insolation $S_{\mathrm{p}}$, and the restricted Jeans escape parameter $\Lambda$ are computed as in \citet{Psaridi2024}. The escape velocity is calculated as $v_{\mathrm{esc}} = (2GM_{\mathrm{P}}/R_{\mathrm{P}})^{1/2}$.
\end{tablenotes}
\label{tab:planetaryparams}
\end{center}
\end{table}

\section{Discussion}\label{sec:Discussion}
\subsection{TOI-1699A b: a well-characterized (S-type) sub-Neptune}\label{sec:TOI-1699A b}
TOI-1699A b adds to the growing population of 176 well-characterized transiting sub-Neptunes from the PlanetS catalog\footnote{Available on the Data \& Analysis Center for Exoplanets (DACE) platform (\url{https://dace.unige.ch}, updated in September 2026).\label{dacefootnote}} ($\sigma_M/M \leq 25\%$ and $\sigma_R/R \leq 8\%$; \citealt{Otegi2020,Parc2024}). TESS is by far the largest single contributor to this sample, having detected 110 of them (63\%). Out of the 176, 61 orbit G-type stars with 31 of them detected by TESS (51\%). TOI-1699A b joins the limited sample of (TESS) sub-Neptunes with precise densities around Sun-like hosts.

\begin{figure*}[t!]
\sidecaption
  \includegraphics[width=12cm]{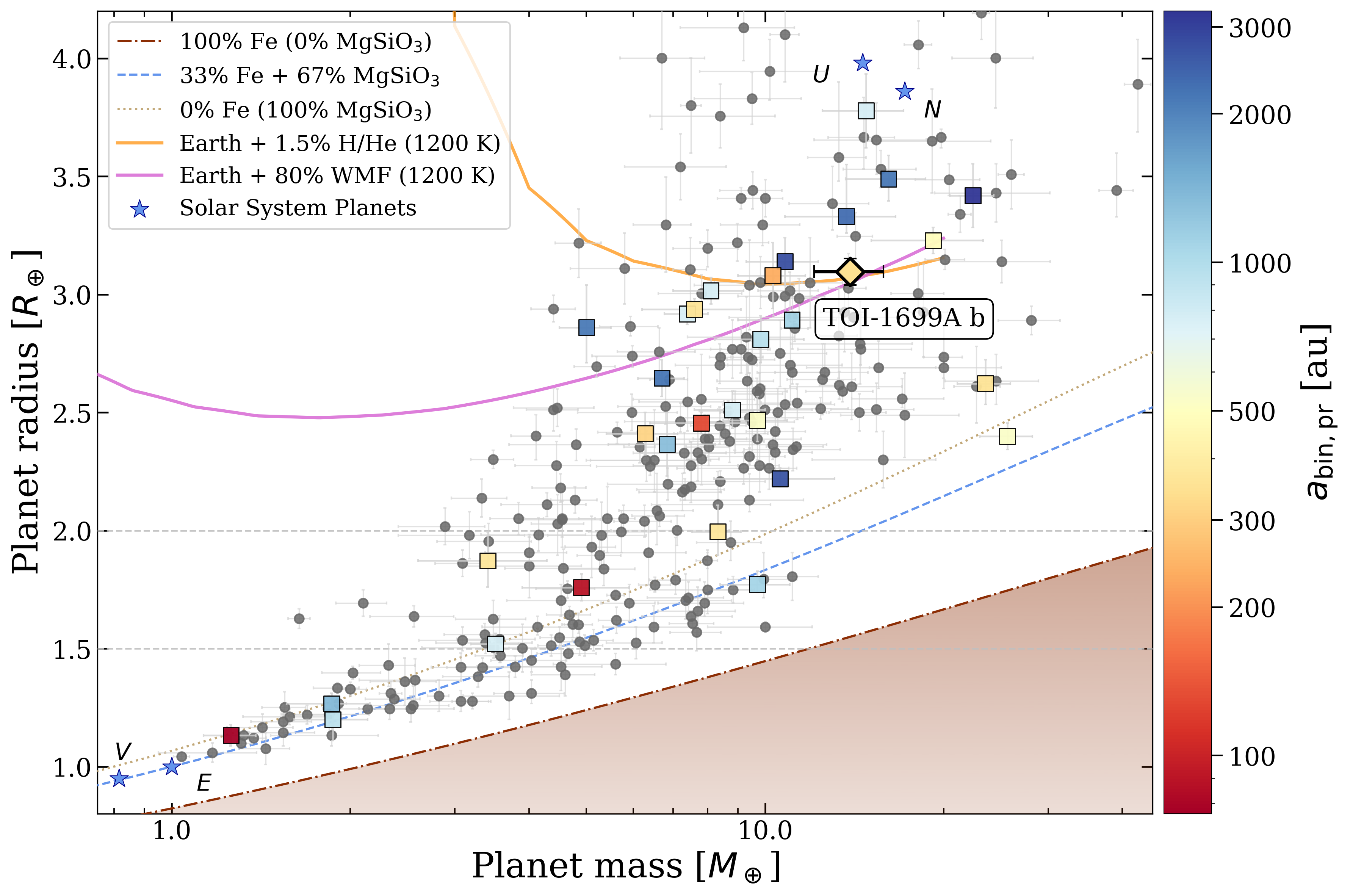}
  \caption{Mass-radius diagram of TOI-1699A b (diamond) and well-characterized planets from the PlanetS catalog (gray). S-type planets are shown as squares color-coded by projected binary separation. The horizontal dashed lines mark the radius valley. Composition curves show pure iron (brown), Earth-like (blue), and pure silicate (beige) interiors \citep{Zeng2019}, an Earth-like interior with a $1.5\%$ H/He envelope (orange; \citealt{Tang2025}), and an Earth-like composition with an $80\%$ water mass fraction (pink; \citealt{Aguichine2025}).}
  \label{fig:massradius}
\end{figure*}

 We place TOI-1699A b on the mass-radius ($M$-$R$) diagram (Fig.~\ref{fig:massradius}), together with the population of well-characterized small (1--4 \re) transiting exoplanets from the PlanetS catalog. We also highlight the 30 S-type small ($\leq$ 4 \re) planets in the sample as color-coded squares, with the color indicating the projected binary separation $a_\mathrm{bin}$ (\citealt{Messamah2026}). With our detection, TOI-1699A b adds to this currently limited sample of well-characterized small S-type planets. The horizontal dashed lines mark the approximate boundaries of the radius valley (1.5--2.0~\re) for single-star planets \citep{Fulton2017, Petigura2022}. With a mass of $13.9^{+1.9}_{-1.8}\,M_\oplus$, a radius of 3.10 $\pm$ 0.06~\re, and a bulk density of 2.6 $\pm$ 0.4~g~cm$^{-3}$, TOI-1699A b lies in the sub-Neptune regime. Its position is well above the theoretical composition curves of rocky compositions, including pure iron (brown dash-dotted line), Earth-like composition (blue dashed line), and pure silicates (beige dotted line) from \citet{Zeng2019}. Therefore, TOI-1699A b cannot be reproduced by a bare rocky interior and requires a significant amount of volatiles to explain its size. In Fig.~\ref{fig:massradius}, we also plot two volatile-rich composition lines that match its equilibrium temperature ($T_{\rm eq}\simeq1200$~K) and age (5.4~Gyr): an Earth-like interior surrounded by a $1.5\%$ H/He envelope (orange; \citealt{Tang2025}) and an Earth-like composition with an $80\%$ water mass fraction (pink; \citealt{Aguichine2025}). TOI-1699A b lies close to both lines, highlighting the strong $M$-$R$ degeneracy of sub-Neptunes, whose bulk densities can be reproduced by either H/He-dominated ``gas dwarfs'' or ``water-rich'' ocean worlds \citep{Parc2024}.

With a radius of 3.10 $\pm$ 0.06~\re, TOI-1699A b lies at the radius cliff, the sharp drop in planet occurrence above $\sim3~\re$ \citep{Fulton2017, Kite2019}. To explain this effect, \citet{Kite2019} suggested that hydrogen dissolves into the magma ocean once the pressure at the base of the atmosphere is high enough, which slows down the rate of growth in planet radius. In addition, \citet{Burn2024} reproduced the radius cliff using a combined planet formation and evolution model for planets with fully mixed H/He and water envelopes. Well-characterized planets that lie on the cliff, like TOI-1699A b, can shed light on this effect.

With an equilibrium temperature of $T_{\rm eq}$ = 1200 $\pm$ 18~K and an insolation of $S_p = 347^{+21}_{-20}\,S_\oplus$, TOI-1699A~b has retained a substantial volatile envelope over its $5.4$~Gyr lifetime. Its escape velocity of 23.7 $\pm$ 1.6~km~s$^{-1}$ is similar to that of Neptune, and the restricted Jeans escape parameter is $\Lambda = 28.3^{+3.9}_{-3.7}$ \citep{Fossati2017}, which excludes that the planet went or is going through a phase of catastrophic mass loss. Nevertheless, the planet likely went through a phase of strong hydrodynamic loss in the past, which may still be ongoing and possibly lead to the escape of a significant fraction of the atmosphere, but detailed mass loss and atmospheric evolution calculations would be required to assess this. In any case, its volatile content should still be informative on how the planet was formed and migrated, which we discuss further in Sect.~\ref{sec:formation}.

Theoretical models predict that the closer the stellar companion is, the stronger the truncation of the protoplanetary disk \citep{Artymowicz1994, Zagaria2021}, which reduces the reservoir of solid and gaseous material available for planet formation \citep{Venturini2026, Nigioni2026}. This could in principle shift the rocky-to-volatile transition to smaller planets in such close binaries. At a projected binary separation of $a_\mathrm{bin} \sim348$~au, the disk truncation effect is expected to be mild for a circular orbit \citep{Zagaria2021}, although it could be stronger if the binary is eccentric (Sect.~\ref{sec:formation}). That is consistent with TOI-1699A~b having retained a substantial volatile envelope and suggests that sufficient disk material was available for planet formation despite the stellar multiplicity. 

\subsection{Interior modeling of TOI-1699A b}\label{sec:interior}
As a next step, we use the open-source \plaNETic framework\footnote{\url{https://github.com/joannegger/plaNETic}} \citep{Egger2024} to infer the internal structure of TOI-1699A b. \plaNETic uses a neural network trained on the \texttt{BICEPS} planetary structure model \citep{Haldemann+2024} as a fast surrogate for the forward model. The planet is modeled as an iron-sulfur core, a silicate mantle, and a fully mixed volatile layer containing H/He and water. We ran six models, considering two options for the composition of the volatile layer (A: water-rich, B: water-poor) and three priors for the planetary Si/Mg/Fe ratios (1: stellar, 2: iron-enriched following \citealt{Adibekyan+2021}, 3: uniform).

\begin{figure*}[t]
    \centering
    \includegraphics[width=\textwidth]{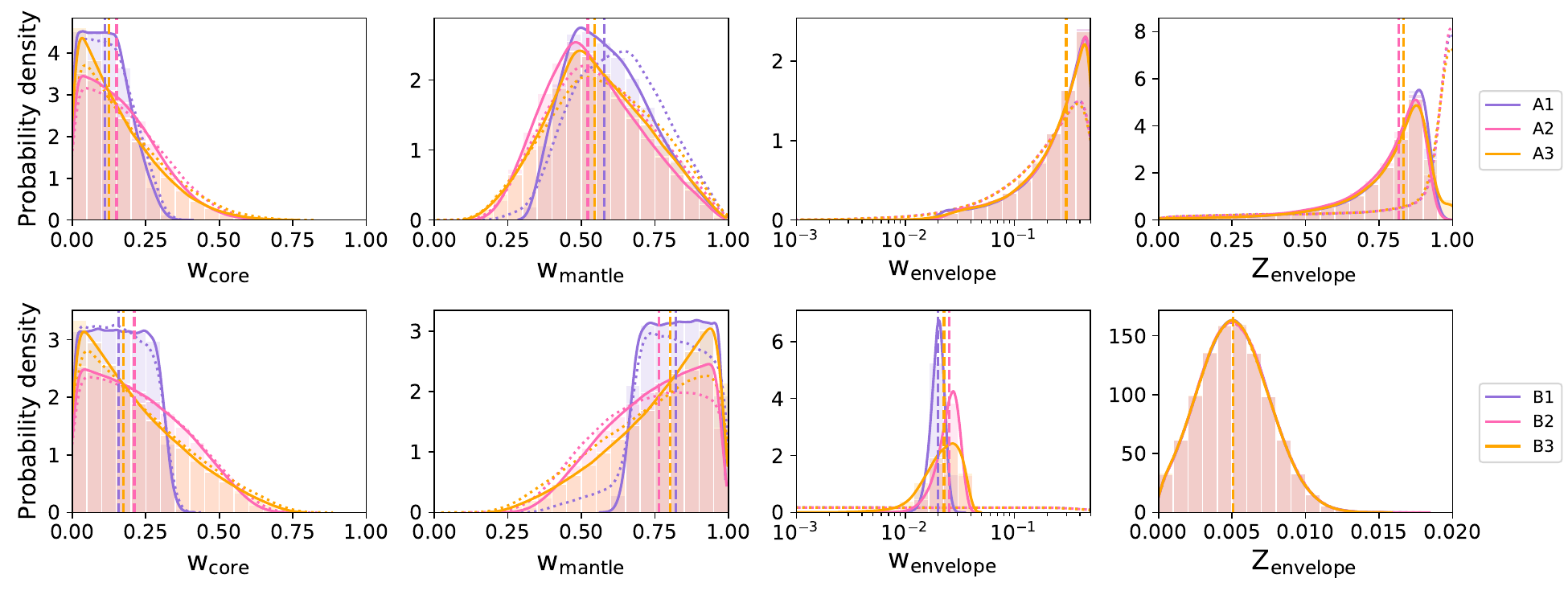}
    \caption{Posteriors of the most important internal structure parameters of TOI-1699A\,b derived with the \texttt{plaNETic} framework, namely the mass fractions of the inner core, mantle, and volatile layers, as well as the mass fraction of water in the volatile layer, denoted by $Z_{\mathrm{envelope}}$. Solid lines show the posteriors, dotted lines the corresponding priors, and dashed vertical lines the median values of the posterior distributions. In the top row, the results for the water-rich prior (option A) are depicted, in the bottom row for the water-poor prior (option B). Colors depict the Si/Mg/Fe priors (stellar, iron-enriched, free).}
    \label{fig:int_struct}
\end{figure*}

The resulting posterior distributions of the most relevant internal structure parameters are shown in Fig. \ref{fig:int_struct}, while a complete summary of the posteriors is provided in Table \ref{tab:internal_structure_results_b}. As expected for a planet with the mass and radius of TOI-1699A b, we are not able to constrain the mass fractions of the inner core and mantle layers; our derived posteriors mostly match the chosen priors, and the results for the three Si/Mg/Fe priors differ only slightly. In the water-rich case, a very large range of envelope mass fractions and water mass fractions can explain the observed planetary parameters, from pure H/He envelopes to almost 100\% water. For the water-poor prior, the inferred envelope mass fractions have median values of around 2\%. However, we point out that this value reflects the adopted water-poor prior rather than a genuine data-driven constraint, since the mass and radius of TOI-1699A b alone cannot break the H/He-water envelope degeneracy.

\subsection{Estimated physical properties of TOI-1699A c}\label{sec:TOI-1699A c}
TOI-1699A c was detected in the HARPS-N RVs as a massive outer companion with an orbital period of $\sim$109 days. As the available TESS observations do not cover the predicted transit windows of the planet, its transiting nature cannot be constrained photometrically. Therefore, our analysis provides only a minimum mass of $M_\mathrm{c}\sin i_\mathrm{c}$ = 5.54 $\pm$ 0.21 M$_\mathrm{Jup}$. The measured minimum mass lies well within the planetary-mass regime, while the true mass of TOI-1699A~c would exceed the deuterium-burning limit of $13\,M_{\rm Jup}$ for orbital inclination lower than $i_c \simeq 25$~deg. Assuming coplanar orbits and using the inclination of the inner planet ($i_\mathrm{b}$ = 86.88 $\pm$ 0.17~deg), we estimated the expected impact parameter of the outer planet using Eq. 7 of \citet{Winn2010}. Using the posterior parameters from the joint fit, we obtain $b_\mathrm{c}$ = 5.3 $\pm$ 0.3. Therefore, under the assumption of coplanarity, TOI-1699A c is not expected to transit. Despite its large mass, the wide period ratio ($P_c/P_b \simeq23.9$) results in negligible transit timing variations (TTVs) on the inner planet. This is supported by dynamical simulations using {\tt TTVFast} (\citealt{Deck2014}), which highlight TTVs of less than 1 minute on planet b for $e_\mathrm{c}$ < 0.5. This is below the timing precision of the TESS and CHEOPS observations. Moreover, under the coplanar assumption, the true mass of TOI-1699A c is expected to be nearly identical to its minimum mass. Using the empirical giant-planet $M$-$R$ relation from \citet{Muller2024}, $\rplanet = (18.6 \pm 6.7)\,\mplanet^{-0.06 \pm 0.07}$ for $\mplanet \gtrsim 127~\me$, we estimate a planetary radius of $R_\mathrm{c} \sim11.9~\re$, consistent with a giant gaseous planet.

\subsection{TOI-1699A among (circumstellar) multiplanet systems}\label{sec:circumstellarmulti}
According to the NASA Exoplanet Archive\footnote{\url{https://exoplanetarchive.ipac.caltech.edu}, updated in September 2026.}, only a handful of systems consist of a well-characterized inner sub-Neptune with a long-period massive companion. Motivated by the similar architecture of TOI-756 reported by \citet{Parc2025}, we searched for systems hosting an inner transiting planet with $2 \leq R_{\mathrm{p}} \leq 4~\re$, $P_{\mathrm{in}} \leq 10$~days, and radius and mass precisions smaller than $8\%$ and $25\%$, respectively, similar to the PlanetS catalog. We then required at least one outer companion with $P_{\mathrm{out}} \geq 100$~days and either $R_{\mathrm{p,out}}>4~\re$ or a measured mass $M_{\mathrm{p,out}}>20~\me$ ($\sim0.06\,M_{\mathrm{Jup}}$; for nontransiting RV planets, the corresponding minimum mass $M_{\mathrm{p,out}}\sin i$ was used). Only ten systems in the NASA archive satisfy these criteria, corresponding to 13 inner-outer planet pairs. Using a more restrictive cut of $M_{\mathrm{p,out}}>0.3~M_{\mathrm{Jup}}$, or $M_{\mathrm{p,out}}\sin i>0.3~M_{\mathrm{Jup}}$ for RV-only companions, the number of pairs is reduced to 12 but leaves the number of systems unchanged. 

We also repeated our search for nontransiting inner planets detected only with RVs. Since no radius is available for these planets, we selected the inner planets using the following criteria: $5 \leq M_{\mathrm{p}}\sin i \leq 20~\me$, $P_{\mathrm{in}}\leq10$~days, and a mass precision better than $25\%$. For the outer companions, we applied the same criteria as above: $P_{\mathrm{out}}\geq100$~days and either $R_{\mathrm{p,out}}>4~\re$ or a measured mass $M_{\mathrm{p,out}}>20~\me$, using $M_{\mathrm{p,out}}\sin i$ for RV-only companions. With these criteria, we find 27 systems, corresponding to 34 inner-outer planet pairs. Ten of these systems are also present in the transiting-inner-planet sample, while 17 additional systems are identified only when nontransiting inner planets are included. Applying the stricter outer-companion cut of $M_{\mathrm{p,out}}>0.3~M_{\mathrm{Jup}}$ (or $M_{\mathrm{p,out}}\sin i>0.3~M_{\mathrm{Jup}}$ for RV-only companions) reduces the sample to 21 systems and 25 pairs. 

The TOI-1699A system therefore adds to this relatively limited sample of planetary systems, consisting of a short-period low-mass planet together with a long-period massive companion. The small but nonzero sample is important to understand how outer giants shape inner small-planet systems. Rather than the inner regions being cleared, as expected if a massive giant had dynamically scattered close-in low-mass planets (\citealt{Juric2008}; \citealt{Bitsch2020}), these systems show that inner sub-Neptunes can survive alongside an outer giant, in line with the positive correlation reported between the two populations (\citealt{ZhuWu2018}; \citealt{Bryan2019}). Our system stands out by hosting one of the most massive companions at one of the shortest outer-planet periods, and yet its inner planet has grown enough to retain a substantial volatile envelope. The low eccentricity of TOI-1699A c ($e_\mathrm{c}$ = 0.074) points to a dynamically quiet history, consistent with an inner planet that was never violently perturbed. TOI-1699A is therefore an addition to the small sample of systems with such architecture.

With the detection of TOI-1699A c, TOI-1699A joins the still limited sample of multiplanet systems in S-type stellar binaries \citep[e.g.,][]{Psaridi2024, Parc2025}. We used the catalog of \citet{Thebault2025}\footnote{\url{https://exoplanet.eu/planets_binary/}, updated in September 2026.} that consists of 850 S-type systems. Among these, 172 host two or more confirmed planets, with the large majority of S-type planetary systems hosting a single detected planet. We acknowledge that this multiplicity fraction is only a lower limit given that stellar companions are often unresolved, and the detection of a second planet requires either a favorable transit geometry or a long-enough RV baseline. The apparent scarcity of multiplanet S-type systems is therefore at least partly observationally biased. Using the \citet{Thebault2025} catalog, we find no significant difference between the binary separation distributions of single-planet and multiplanet S-type systems ($p = 0.76$; Kolmogorov--Smirnov test), with median separations of $\sim$980 and $\sim$1030~au, respectively. TOI-1699, with a projected separation of $\sim$348~au, lies on the lower side of this distribution. We note that the separation distribution of planet-hosting binary systems is strongly biased, since close pairs are avoided by RV surveys while unresolved companions could be completely missed. Finally, for multiple-star systems with planets within 100\,pc, \citet{GonzalezPayo2024} found no significant difference in the number of planets between single and multiple-star systems, nor any dependence on the ratio between the binary separation and the semimajor axis of the outermost planet ($s/a$). Planets in multiple-star systems at small $s/a$ seem to have larger eccentricities than those around single stars, but there is no significant trend with the binary separation alone.

\subsection{Formation of TOI-1699A b and c}\label{sec:formation}
The stellar companion to TOI-1699A is observed at an angular separation of 1.68~arcsec. Taking the observed distance to the system (206.9 pc), and applying the statistical correction of \citet{DuquennoyMayor1991}, we estimate a physical binary separation ($a_{\rm bin}$) of approximately 470 au, which we emphasize is not necessarily the true binary separation. Using the disk truncation prescription of \cite{Venturini2026} and the stellar masses, a binary separation of 470 au corresponds to a truncation radius of approximately 187 au for a circular binary orbit. In the absence of constraints on the binary eccentricity, this estimate represents an upper limit. If the stellar companion follows an eccentric orbit, the circumprimary disk could have experienced substantially stronger truncation, reducing the available solid and gaseous reservoir for planet formation. 

To investigate possible formation pathways, we searched for analogs in the synthetic S-type population of \citet{PAIRSIII}, following an approach similar to \citet{Ulmer-Moll2023} and \citet{Egger2024}. For TOI-1699A\,b, we selected systems within 25\% of the observed $\log a$, $\log R_\mathrm{P}$, and $\log M_\mathrm{P}$; for TOI-1699A\,c, systems within 25\% in $\log a$ and with a mass above the observed value (allowing a 25\% tolerance on this lower limit). Among these, we kept the analog whose binary separation is closest to the approximate estimate of $a_\mathrm{bin} \sim 470$\,au.

This criterion yielded one analog of TOI-1699A b. This analog formed in a binary with a separation of $a_{\rm bin}=466.6$ au. Furthermore, the final planetary eccentricity of the analog is $e=0.16$, consistent with the observational constraint of $e<0.27$ (3$\sigma$ upper limit reported in Table \ref{tab:planetaryparams}). This planet has $\sim$42\% of its total mass made of water and $\sim$5\% made of H/He. Such high water fraction indicates an origin beyond the water ice line. The analog subsequently migrated inward to its present location while accreting a gaseous envelope from the protoplanetary disk. 

In contrast, no analog was found for planet c. This discrepancy may reflect limitations of the current population synthesis setup rather than an intrinsic difficulty in forming such an object. Nevertheless, we highlight that because the true orbit of the TOI-1699 binary is unknown, performing a direct comparison with synthetic models is challenging. Further modeling is needed to properly constrain how this system formed, including testing other formation scenarios such as early disk instability that could have led to the formation of the outer giant \citep{Schib2025, Zhang2026}. 

\subsection{Sensitivity to hidden planets}\label{sec:ardent}
The discovery of TOI-1699A c poses important questions about the system formation. The exact impact that such outer massive companions play on inner super-Earths and sub-Neptunes is under active debate \citep[e.g.,][ and references therein]{Bryan2024, Bonomo2025}. Constraining the architecture of the inner systems provides useful insight once compared with systems that do not contain outer giants. To this end, we computed planet detection limits which account both for the RV data, and the dynamical influence of the planets. For this analysis, we used the Algorithm for the Refinement of DEtection limits via N-body stability Threshold (\ARDENT, \citealt{Stalport2025})\footnote{\url{https://github.com/manustalport/ardent}}. 

\begin{figure}[t!]
  \centering
  \includegraphics[width=.50\textwidth]{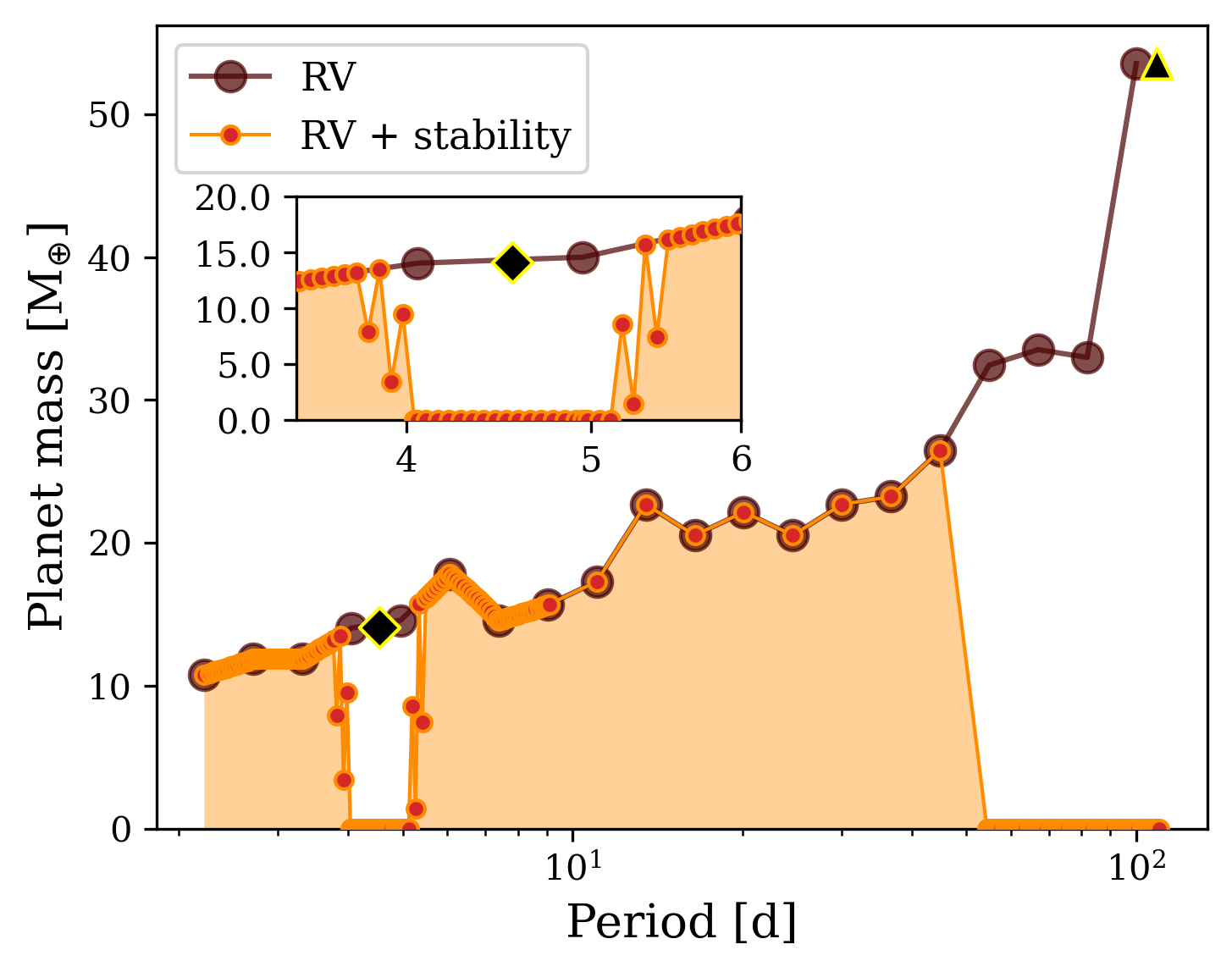}
\caption{Planet detection limits in the inner system of TOI-1699A computed with \ARDENT. The data-driven detection limits computed via injection-recovery tests are shown in brown. The dynamical detection limits are in orange. TOI-1699A b and TOI-1699A c are illustrated with black symbols (the upward triangle for TOI-1699A c means the planet mass lies above the represented mass range). The inset plot is a zoom-in onto planet b. The zone where additional planets could still be found is highlighted in orange.}
  \label{fig:DetectLimits}
\end{figure}

We first performed planet injection-recovery tests in the inner system, between 2 and 110 days. We randomly generated 10$^4$ points in the $P$-$K$ space, following a log-uniform distribution along the orbital period. Regarding $K$, we employed a uniform distribution between 0.5 and 3 times the residual RV rms. For each point, we performed ten injection-recovery tests in the RV residuals with evenly spaced orbital phases, and null eccentricity. From this ensemble of tests, we computed the 95$\%$ detection limits. This is represented by the brown curve in Fig. \ref{fig:DetectLimits}. These data-driven mass limits increase sharply near the upper end of the period range, as an effect of the limited RV time span (172 days) which decreases the sensitivity to signal detection beyond $\sim$100 days.
TOI-1699A b stands on the detection limit, which does not account for the photometric information. 

Secondly, we investigated the dynamical plausibility of the injected planets under the gravitational influence of TOI-1699A\,b and c. We retrieved their planet parameters from Table \ref{tab:planetaryparams} and assumed coplanar orbits. \ARDENT combines analytical and numerical stability criteria to further exclude zones of the $P$-$M$ space that are dynamically unstable. The orange curve in Fig. \ref{fig:DetectLimits} illustrates the dynamical detection limits computed with default settings (integration time set to $10^5\times P_c$, cf. \citealt{Stalport2025}). Inside the shaded area, the injected planets remain undetected and are dynamically plausible. They show that the gravitational influence of TOI-1699A c prevents the existence of any additional planet between 54 and 110~days. As a result, we can also discard the presence of additional objects more massive than 27~\me\ interior to planet c. Furthermore, these results highlight the (smaller) dynamical influence of the inner planet, TOI-1699A b. It cleans out a zone between 4 and 5.1 days, as is better observed in the inset plot. Adjacent to this forbidden period range, the sawtooth structures are indicative of mean-motion resonance overlaps (e.g., \citealt{Stalport2025}). Finally, we carried out a similar analysis extending up to 500 days. While the data-driven detection limits are saturated and meaningless beyond $\sim$100 days, the dynamical constraints allow us to exclude additional unseen planets between 54 and 235 days.

\subsection{Stability and future prospects}\label{sec:future}

We further explored the dynamical stability of the system as a function of the unknown inclination, $i_\mathrm{c}$, and longitude of ascending node, $\Omega_\mathrm{c}$, of TOI-1699A~c, following the approach of previous dynamical studies of multiplanet systems \citep{Correia2005,Correia2010}. We assumed that the stellar companion TOI-1699B follows a circular orbit coplanar with TOI-1699A~b, while keeping the remaining orbital parameters fixed at their best-fit values in Table~\ref{tab:planetaryparams}, including the RV semi-amplitude $K_c$. The system was integrated with the SABAC4 symplectic integrator \citep{LaskarRobutel2001}, including general relativistic corrections, and stability was assessed through frequency analysis \citep{Laskar1990,Laskar1993}, using the diffusion of the mean motion of TOI-1699A~b as a stability indicator.

The nominal prograde coplanar configuration lies well inside a broad stable region, while the stability structure is mainly controlled by the mutual inclination between planets b and c. Broad stable domains are found for $i_{\mathrm{mut}}\lesssim70$~deg and $i_{\mathrm{mut}}\gtrsim110$~deg, whereas most near-polar configurations are dynamically unstable. Nearly face-on solutions, with $i_\mathrm{c}\lesssim20$~deg or $i_\mathrm{c}\gtrsim160$~deg, are also dynamically disfavored. Since $K_\mathrm{c}$ is kept fixed, the true mass of the outer companion increases as the orbital inclination decreases. Combined with the measured $M_\mathrm{c}\sin i_\mathrm{c}=5.54 \pm 0.21~M_{\mathrm{Jup}}$, the stability constraint therefore translates into an approximate, model-dependent upper limit of $M_\mathrm{c}\lesssim18~M_{\mathrm{Jup}}$. Dynamical stability thus disfavors a stellar-mass nature for TOI-1699A~c, although a low-mass brown-dwarf configuration remains possible. This constraint is complementary to future \textit{Gaia} DR4 astrometry combined with continued RV monitoring, which may further constrain the inclination and true mass of the outer companion.

For an edge-on orbit with $M_\mathrm{c}=5.54~M_\mathrm{Jup}$, the astrometric reflex signature of TOI-1699A would be only $\upsilon=0.012$\,mas, far below the nominal per-epoch along-scan precision of $\sigma_\mathrm{AL} = 0.034$\,mas; a signal-to-noise ratio of $\upsilon/\sigma_\mathrm{AL}\gtrsim2$ is required for a robust detection in a blind search \citep[e.g.,][]{Ranalli2018}. However, the combination of astrometry and RVs can meaningfully constrain orbital inclinations, and hence true masses, even when the astrometric signal by itself is undetectable \citep{Espinoza-Retamal2023}. MCMC fits of our RV data together with synthetic \emph{Gaia} DR4 data suggest that for an edge-on orbit, the orbital inclination will be recovered with a precision of 13~deg, while for an inclination of 30~deg the precision will be 6~deg, distinguishing the cases of TOI-1699A~c being a planet or a brown dwarf. The \emph{Gaia} visit epochs and scan angles were simulated with the \texttt{gaiascanlaw} tool \citep{Guerriero2026}\footnote{Available at \url{https://github.com/zpenoyre/gaiascanlaw}.}. A complicating factor however is that both TOI-1699A and B show an elevated astrometric noise, each having a renormalized unit weight error (RUWE) of 1.46. For TOI-1699A this corresponds to an astrometric excess noise of $0.2$\,mas which cannot be attributed to either the binary orbit or the effects of a planet-mass companion TOI-1699A~c; instead, this likely arises from blending contaminating the astrometric measurements \citep{ElBadry2021,Fabricius2021}. The prospects for a mass and inclination measurement for planet c will depend on the extent to which this can be corrected for.

\section{Conclusions}\label{sec:Conclusions}
We presented the detection and characterization of the TOI-1699A multiplanet system. The system consists of TOI-1699A b, a transiting sub-Neptune, and TOI-1699A c, a massive nontransiting outer planet. Both planets orbit the G5\,V primary of a wide stellar binary with an M3\,V companion at a projected separation of $\sim$348~au, forming an S-type planetary system. TOI-1699A b was initially detected by TESS and later observed with targeted CHEOPS photometry, ground-based multiband photometry, and HARPS-N RVs. TOI-1699A c was discovered with our HARPS-N observations. The negligible contamination of the HARPS-N fiber by the faint secondary star indicates that TOI-1699A c also orbits the primary star.

The inner planet is a sub-Neptune with an orbital period of 4.55~days, a radius of 3.10 $\pm$ 0.06~\re, and a mass of $13.9^{+1.9}_{-1.8}~\me$. Its bulk density of $2.6 \pm 0.4$~g~cm$^{-3}$ cannot be explained by a bare rocky interior. Our internal structure modeling results in a composition that is consistent with either a water-rich envelope with a broad range of envelope mass fractions or, using a water-poor prior, a H/He envelope of $\sim$2\% by mass. Our results reflect the compositional degeneracy of 2--4~\re sub-Neptunes. The outer planet, TOI-1699A c, has an orbital period of 108.75 $\pm$ 0.10~days, an eccentricity of 0.074 $\pm$ 0.003, and a minimum mass of 5.54 $\pm$ 0.21~$M_{\mathrm{Jup}}$. Assuming coplanarity with the inner planet, it is not expected to transit ($b_{\mathrm{c}} \approx 5.3$), and its true mass would be close to the minimum one.

Our combined data-driven and dynamical detection-limit analysis with \ARDENT\ shows that, due to the gravitational influence of TOI-1699A c, no additional planets can exist between 54 and 235~days, while at shorter periods we exclude planets more massive than 27~\me. Moreover, our dynamical stability analysis sets an upper mass limit of $M_{\mathrm{c}} \lesssim 18~M_{\mathrm{Jup}}$ for TOI-1699A~c, disfavoring a stellar-mass nature.

TOI-1699A adds to the still limited sample of S-type multiplanet systems, showing that an inner sub-Neptune can retain a substantial volatile envelope alongside a nearby massive companion. Our comparison with a synthetic S-type population reproduces TOI-1699A b with an analog that formed beyond the water ice line and migrated inward. However, we do not find an exact two-planet analog, highlighting the need for further population-synthesis development. The upcoming \textit{Gaia} DR4 astrometry and RV monitoring will help pin down the true mass and inclination of TOI-1699A c, further establishing TOI-1699A as a valuable benchmark for planet formation in S-type systems.

\section*{Data availability}
The CHEOPS photometry and the HARPS-N RVs in this work are available at the CDS via anonymous ftp to \url{cdsarc.cds.unistra.fr} or via \url{https://cdsarc.cds.unistra.fr/viz-bin/cat/J/A+A/}.

\begin{acknowledgements}
This study uses \textit{TESS} data, CHEOPS data observed as part of guaranteed time observation program CH\_PR140085 (PI: Venturini), and HARPS-N data under program CAT25B\_53 (PI: Psaridi).

A.P., R.A., D.B., E.P., G.L., J.K., F.M., I.R., and E.V. acknowledge financial support from MICIU/AEI/10.13039/501100011033 and the ERDF ``A way of making Europe'' through projects PID2021-125627OB-C31, PID2021-125627OB-C32, PID2021-127289NB-I00, PID2023-150468NB-I00, PID2023-149439NB-C41, PID2024-158486OB-C31, and PID2024-158486OB-C32; from the ``Severo Ochoa'' Centre of Excellence award to the Instituto de Astrof\'isica de Canarias (CEX2019-000920-S) and the ``Mar\'ia de Maeztu'' Unit of Excellence award to the Institut de Ci\`encies de l'Espai (CEX2020-001058-M); and from the Generalitat de Catalunya/CERCA programme.

A.P. additionally acknowledges support from the MaX-CSIC Excellence Award MaX4-SOMMA-ICE, from the MICIU and the AEI through the Juan de la Cierva programme (contract JDC2024-054476-I), and from the European Research Council (ERC) under the European Union's Horizon Europe programme (ERC Advanced Grant SPOTLESS; no.~101140786).

G.L., F.M., and E.P. acknowledge support from the ERC (ERC AdvG SPEAR, GA 101200674).

N.C.S. acknowledges funding by the European Union (ERC, FIERCE, 101052347).

The ERC projects SPOTLESS, SPEAR, and FIERCE are funded by the European Union. Views and opinions expressed are however those of the author(s) only and do not necessarily reflect those of the European Union, the European Research Council, or the European Research Council Executive Agency (ERCEA). Neither the European Union nor the granting authority can be held responsible for them.

J.V., A.D., A.L., M.L., and B.A. acknowledge support from the Swiss National Science Foundation (SNSF): J.V. under grant PZ00P2\_208945, A.D. under grant 200021\_200726, A.L. under grant TMSGI2\_211697, and B.A. and M.L. under grant PCEFP2\_194576. B.A. additionally acknowledges support from the Australian Government through the Australian Research Council's Discovery Projects funding scheme (project DP250103019). D.E. acknowledges financial support from the SNSF.

Several authors acknowledge support from Funda\c{c}\~ao para a Ci\^encia e Tecnologia (FCT), Portugal: S.G.S. through contract CEECIND/00826/2018 and POPH/FSE (EC); A.C.M.C. through the CFisUC project UID/04564/2025 (DOI 10.54499/UID/04564/2025); V.A. through the FCT Scientific Employment Stimulus program (reference 2023.06055.CEECIND/CP2839/CT0005, DOI 10.54499/2023.06055.CEECIND/CP2839/CT0005); S.C.C.B. through the Scientific Employment Incentive program (reference 2023.06687.CEECIND, DOI 10.54499/2023.06687.CEECIND/CP2839/CT0002); and O.D.S.D. through work contract DL 57/2016/CP1364/CT0004, funded by national funds through FCT.

A.C.M.C., A.D., B.E., K.G., and J.K. acknowledge their roles as ESA-appointed CHEOPS Science Team Members.

K.A.C. and E.N. acknowledge support from the TESS mission via NASA grants 80NSSC24K1889 and 80NSSC26K0081; K.A.C. additionally via subaward s3449 from MIT.

A.B., M.F., and A.J.M. acknowledge support from the Swedish National Space Agency (SNSA): M.F. under DNR 65/19 and 174/18, and A.J.M. under career grant 2023-00146.

M.G. is an F.R.S.-FNRS Research Director and V.V.G. is an F.R.S.-FNRS Research Associate.

D.G. acknowledges financial support from the CRT Foundation under Grant No.~2018.2323, ``Gaseous or rocky? Unveiling the nature of small worlds''.

J.A.E. acknowledges support through the European Space Agency (ESA) Research Fellowship programme in Space Science.

T.G.W. acknowledges support from the UKSA and the University of Warwick. A.C.C. acknowledges support from STFC consolidated grant number ST/V000861/1 and UKSA grant number ST/X002217/1. P.F.L.M. acknowledges support from STFC research grant number ST/R000638/1. N.A.W. acknowledges UKSA grant ST/R004838/1.

L.B., V.N., I.P., G.P., R.R., and G.S. acknowledge support from CHEOPS ASI-INAF agreement n.~2019-29-HH.0.

C.B., A.Peel, and A.S. acknowledge support from the Swiss Space Office through the ESA PRODEX program.

P.E.C. is funded by the Austrian Science Fund (FWF) Erwin Schr\"odinger Fellowship, program J4595-N.

B.-O.D. acknowledges support from the Swiss State Secretariat for Education, Research and Innovation (SERI) under contract number MB22.00046.

M.N.G. is the ESA CHEOPS Project Scientist and Mission Representative. B.M. is the ESA CHEOPS Project Scientist. K.G.I. was the ESA CHEOPS Project Scientist until the end of December 2022 and Mission Representative until the end of January 2023. All of them are/were responsible for the Guest Observers (GO) Programme. None of them relay/relayed proprietary information between the GO and Guaranteed Time Observation (GTO) Programmes, nor do/did they decide on the definition and target selection of the GTO Programme.

K.W.F.L. acknowledges support from Deutsche Forschungsgemeinschaft grant RA714/14-1 within the DFG Schwerpunkt SPP~1992, Exploring the Diversity of Extrasolar Planets.

D.Q. acknowledges that this work was partially supported by a grant from the Simons Foundation (PI: Queloz, grant number 327127).

G.M.S. acknowledges the SNN-147362 and ADVANCED-153410 grants of the National Research, Development and Innovation Office (NKFIH, Hungary), and the ESA PRODEX Experiment Agreements No.~4000137122 and No.~4000149203.

CHEOPS is an ESA mission in partnership with Switzerland with important contributions to the payload and the ground segment from Austria, Belgium, France, Germany, Hungary, Italy, Portugal, Spain, Sweden, and the United Kingdom. The CHEOPS Consortium would like to gratefully acknowledge the support received by all the agencies, offices, universities, and industries involved. Their flexibility and willingness to explore new approaches were essential to the success of this mission. CHEOPS data analyzed in this article will be made available in the CHEOPS mission archive (\url{https://cheops.unige.ch/archive_browser/}).

Funding for the TESS mission is provided by NASA's Science Mission Directorate. This paper includes data collected by the TESS mission that are publicly available from the Mikulski Archive for Space Telescopes (MAST). We acknowledge the use of public TESS data from pipelines at the TESS Science Office and at the TESS Science Processing Operations Center. Resources supporting this work were provided by the NASA High-End Computing (HEC) Program through the NASA Advanced Supercomputing (NAS) Division at Ames Research Center for the production of the SPOC data products.

This work makes use of observations from the LCOGT network. Part of the LCOGT telescope time was granted by NOIRLab through the Mid-Scale Innovations Program (MSIP). MSIP is funded by NSF.

This paper is based on observations made with the MuSCAT3 instrument, developed by the Astrobiology Center (ABC) in Japan, the University of Tokyo, and Las Cumbres Observatory (LCOGT). MuSCAT3 was developed with financial support by JSPS KAKENHI (JP18H05439) and JST PRESTO (JPMJPR1775), and is located at the Faulkes Telescope North on Maui, HI (USA), operated by LCOGT.

This article is based on observations made with the MuSCAT2 instrument, developed by ABC, at Telescopio Carlos S\'anchez operated on the island of Tenerife by the IAC in the Spanish Observatorio del Teide. This work is partly supported by JSPS KAKENHI Grant Numbers JP24H00017, JP25K24620, JP26H01402, JP26K00755, and JP24K00689.

Some of the observations in the paper made use of the High-Resolution Imaging instrument ``Alopeke''. ``Alopeke'' was funded by the NASA Exoplanet Exploration Program and built at the NASA Ames Research Center by Steve B. Howell, Nic Scott, Elliott P. Horch, and Emmett Quigley. Data were reduced using a software pipeline originally written by Elliott Horch and Mark Everett. ``Alopeke'' was mounted on the Gemini North telescope of the international Gemini Observatory, a program of NSF's OIR Lab, which is managed by the Association of Universities for Research in Astronomy (AURA) under a cooperative agreement with the National Science Foundation, on behalf of the Gemini partnership: the National Science Foundation (United States), National Research Council (Canada), Agencia Nacional de Investigaci\'on y Desarrollo (Chile), Ministerio de Ciencia, Tecnolog\'ia e Innovaci\'on (Argentina), Minist\'erio da Ci\^encia, Tecnologia, Inova\c{c}\~oes e Comunica\c{c}\~oes (Brazil), and Korea Astronomy and Space Science Institute (Republic of Korea).

This research has made use of the Exoplanet Follow-up Observation Program website, which is operated by the California Institute of Technology, under contract with the National Aeronautics and Space Administration under the Exoplanet Exploration Program.

\end{acknowledgements}

\bibliographystyle{aa}
\typeout{}
\bibliography{bib}
\appendix
\onecolumn
\section{TESS photometry}
\begin{figure}[!htbp]
  \centering
  \includegraphics[width=1.0\textwidth]{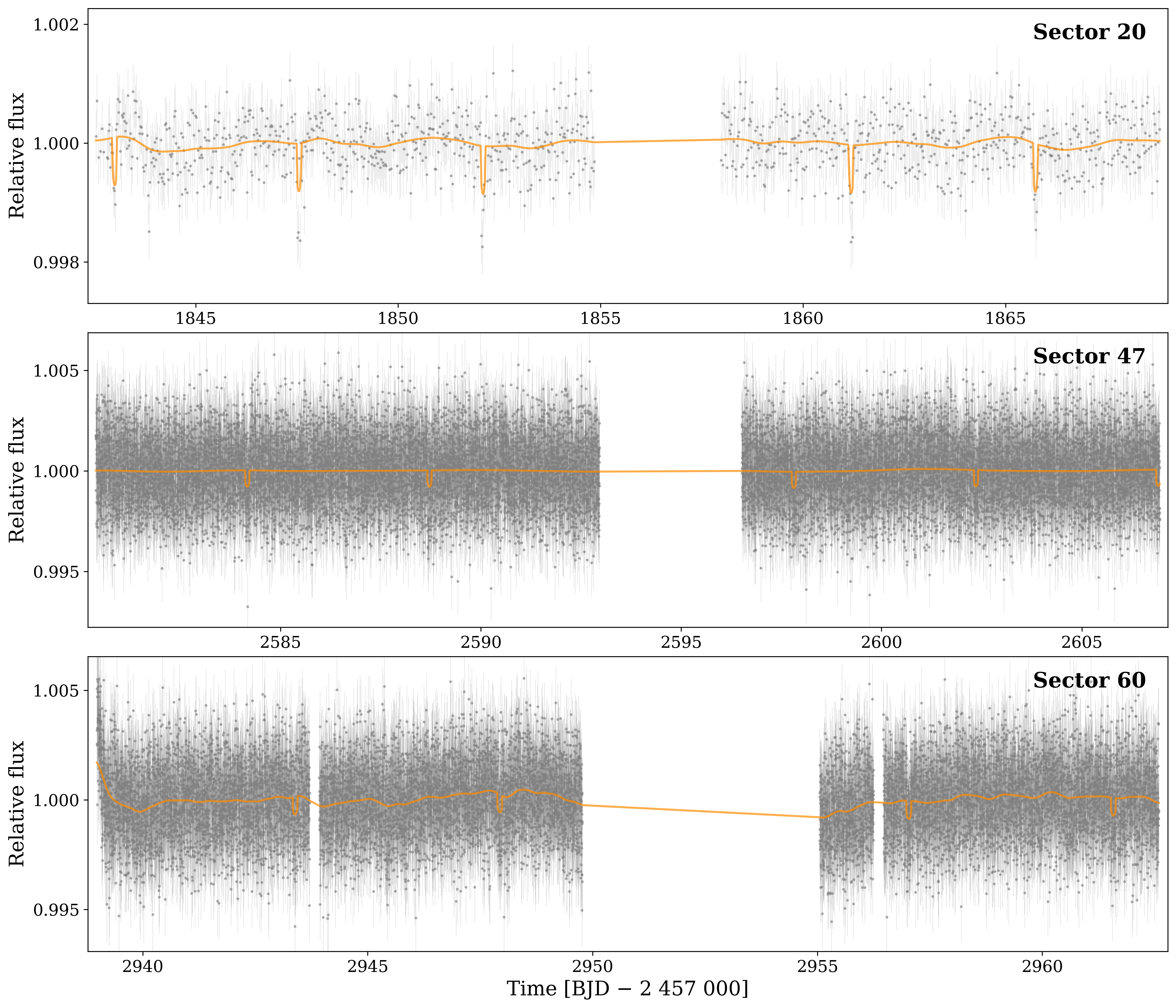}
\caption{TESS PDCSAP light curves of TOI-1699 from Sectors 20, 47, and 60. 
The orange line shows the best-fit transit-plus-GP model of TOI-1699A b 
from the joint analysis (Sect.~\ref{sec:join}).}
  \label{fig:TESS_full}
\end{figure}

\clearpage
\section{HARPS-N radial velocities}
\begin{table*}[!htbp]
\caption{HARPS-N radial velocity measurements and corresponding activity indicators.}
\label{tab:rvs}
\centering
\begin{tabular}{c c c c c c c c c}
\hline\hline
Time & RV & $\sigma_{\mathrm{RV}}$ & FWHM & $\sigma_{\mathrm{FWHM}}$ & Contrast & $\sigma_{\mathrm{Contrast}}$ & BIS & $\sigma_{\mathrm{BIS}}$ \\
{[BJD $-$ 2\,400\,000]} & [m\,s$^{-1}$] & [m\,s$^{-1}$] & [m\,s$^{-1}$] & [m\,s$^{-1}$] & [\%] & [\%] & [m\,s$^{-1}$] & [m\,s$^{-1}$] \\
\hline
60956.72352 & $-54407.6$ & 2.3 & 7083.4 & 4.5 & 60.26 & 0.04 & $-78.8$ & 4.5 \\
60957.72382 & $-54426.5$ & 2.0 & 7096.5 & 4.0 & 60.34 & 0.03 & $-81.0$ & 4.0 \\
60967.72592 & $-54541.4$ & 2.3 & 7079.3 & 4.6 & 60.47 & 0.04 & $-84.4$ & 4.6 \\
60981.72747 & $-54632.9$ & 4.0 & 7084.0 & 7.9 & 60.49 & 0.07 & $-85.4$ & 7.9 \\
60982.71377 & $-54624.7$ & 1.9 & 7092.8 & 3.8 & 60.50 & 0.03 & $-75.3$ & 3.8 \\
60996.64214 & $-54593.1$ & 2.2 & 7086.8 & 4.5 & 60.44 & 0.04 & $-81.6$ & 4.5 \\
60997.67758 & $-54583.1$ & 2.1 & 7091.4 & 4.2 & 60.47 & 0.04 & $-68.8$ & 4.2 \\
61008.58879 & $-54461.5$ & 3.1 & 7086.0 & 6.2 & 60.30 & 0.05 & $-71.0$ & 6.2 \\
61010.68612 & $-54418.7$ & 3.1 & 7104.4 & 6.1 & 60.43 & 0.05 & $-71.0$ & 6.1 \\
61011.72959 & $-54411.0$ & 2.9 & 7098.9 & 5.8 & 60.38 & 0.05 & $-73.5$ & 5.8 \\
61015.66419 & $-54342.2$ & 2.2 & 7094.6 & 4.5 & 60.19 & 0.04 & $-76.3$ & 4.5 \\
61016.77579 & $-54334.4$ & 2.5 & 7087.5 & 5.0 & 60.03 & 0.04 & $-65.7$ & 5.0 \\
61027.69642 & $-54196.2$ & 2.7 & 7068.6 & 5.4 & 60.45 & 0.05 & $-68.8$ & 5.4 \\
61035.64360 & $-54160.5$ & 2.0 & 7100.8 & 4.0 & 60.36 & 0.03 & $-77.7$ & 4.0 \\
61063.42923 & $-54395.8$ & 3.8 & 7082.5 & 7.5 & 60.42 & 0.06 & $-75.4$ & 7.5 \\
61067.55420 & $-54438.8$ & 2.4 & 7090.4 & 4.8 & 60.36 & 0.04 & $-72.1$ & 4.8 \\
61068.57034 & $-54448.6$ & 2.5 & 7096.3 & 5.0 & 60.34 & 0.04 & $-69.8$ & 5.0 \\
61069.53237 & $-54453.2$ & 2.1 & 7091.4 & 4.1 & 60.29 & 0.03 & $-67.0$ & 4.1 \\
61079.56428 & $-54555.7$ & 2.3 & 7087.0 & 4.5 & 60.43 & 0.04 & $-68.3$ & 4.5 \\
61081.51294 & $-54583.5$ & 1.8 & 7096.7 & 3.6 & 60.42 & 0.03 & $-72.7$ & 3.6 \\
61083.61925 & $-54587.1$ & 2.0 & 7088.8 & 4.0 & 60.43 & 0.03 & $-75.0$ & 4.0 \\
61087.48478 & $-54607.5$ & 3.0 & 7086.5 & 6.0 & 60.44 & 0.05 & $-72.9$ & 6.0 \\
61090.50189 & $-54629.3$ & 2.4 & 7077.6 & 4.7 & 60.42 & 0.04 & $-69.8$ & 4.7 \\
61108.47215 & $-54570.8$ & 3.4 & 7093.7 & 6.7 & 60.45 & 0.06 & $-69.9$ & 6.7 \\
61109.46539 & $-54552.3$ & 2.7 & 7095.2 & 5.3 & 60.34 & 0.05 & $-82.4$ & 5.3 \\
61128.43515 & $-54285.1$ & 2.7 & 7101.3 & 5.5 & 60.23 & 0.05 & $-82.2$ & 5.5 \\
\hline
\end{tabular}
\end{table*}
\clearpage
\section{\CONAN joint-fit parameters}
\begin{table*}[!htbp]
\caption{Median values and 68$\%$ confidence intervals of the posterior distributions from joint modeling with \CONAN.}
\small
\centering
\renewcommand{\arraystretch}{1.3}
\setlength{\tabcolsep}{38pt}
\begin{center}
\begin{tabular}{lcc}
\hline\hline
Parameter & Prior$^{(a)}$& Value$^{(b)}$ \\
\hline
Jump stellar parameters & & \\
Stellar density, $\rho_*$ (g~cm$^{-3}$) & $\mathcal{N}(1.24, 0.08)$ & $1.24^{+0.06}_{-0.06}$ \\
Jump parameters for TOI-1699A b & & \\
Orbital period, $\it{P_{orb}}$ (days) & $\mathcal{N}(4.5, 0.1)$ & $4.5470531^{+0.0000052}_{-0.0000052}$ \\
Time of conjunction, $T_\mathrm{c}$ (BJD-2457000) & $\mathcal{N}(2606.9, 0.1)$ & $2606.9075^{+0.0012}_{-0.0013}$ \\
Scaled planetary radius, $R_{P}$/\rstar & $\mathcal{N}(0.03, 0.01)$ & $0.0273^{+0.0004}_{-0.0004}$ \\
Impact parameter, $\textit{b}$ & $\mathcal{N}(0.6, 0.3)$ & $0.603^{+0.024}_{-0.024}$ \\
Eccentricity, $e$ & $\mathcal{F}(0)$ & 0 ($<$ 0.27, 3$\sigma$)$^{(c)}$\\
Argument of periastron, $\omega_*$ (deg) & $\mathcal{F}(90)$ & ... \\
RV semi-amplitude, $\textit{K}$ (m\,s$^{-1}$) & $\mathcal{U}(1, 15)$ & $5.4^{+0.7}_{-0.7}$ \\
Jump parameters for TOI-1699A c & & \\
Orbital period, $\it{P_{orb}}$ (days) & $\mathcal{U}(90, 120)$ & $108.75^{+0.10}_{-0.10}$ \\
Time of conjunction, $T_\mathrm{c}$ (BJD-2457000) & $\mathcal{N}(2868, 20)$ & $2867.8^{+1.1}_{-1.1}$ \\
Eccentricity, $e$ & $\mathcal{U}(0, 1)$ & $0.0737^{+0.0034}_{-0.0034}$ \\
Argument of periastron, $\omega_*$ (deg) & $\mathcal{U}(0, 360)$ & $307.1^{+2.8}_{-2.6}$ \\
RV semi-amplitude, $\textit{K}$ (m\,s$^{-1}$) & $\mathcal{U}(10 , 800)$ & $238.1^{+1.0}_{-1.0}$ \\
GP parameters & & \\
$\sigma_{GP,\mathrm{TESS, S.20}}$ (ppm) & $\mathcal{U}(0, 1000)$ & $90^{+23}_{-27}$ \\
$\rho_{GP,\mathrm{TESS, S.20}}$ (days) & $\mathcal{U}(0, 15)$ & $0.5^{+0.5}_{-0.2}$ \\
$\sigma_{GP,\mathrm{TESS, S.47}}$ (ppm) & $\mathcal{U}(0, 1000)$ & $58^{+31}_{-39}$ \\
$\rho_{GP,\mathrm{TESS, S.47}}$ (days) & $\mathcal{U}(0, 15)$ & $1^{+7}_{-1}$ \\
$\sigma_{GP,\mathrm{TESS, S.60}}$ (ppm) & $\mathcal{U}(0, 1000)$ & $430^{+238}_{-90}$ \\
$\rho_{GP,\mathrm{TESS, S.60}}$ (days) & $\mathcal{U}(0, 15)$ & $0.5^{+0.6}_{-0.2}$ \\
Instrumental RV parameters & & \\
Jitter, $\sigma_{w,\mathrm{HARPS-N}}$ ($\ms$) & $\mathcal{U}(0, 25)$ & $1.3^{+0.8}_{-0.7}$ \\
Systemic RV, $\mu_{\mathrm{HARPS-N}}$ ($\ms$) &$\mathcal{N}(-54400, 300)$  & $-54404.2^{+0.7}_{-0.7}$ \\
Limb darkening parameters & & \\
$q_{1,\mathrm{TESS}}$ & $\mathcal{N}(0.3141, 0.0095)$ & $0.312^{+0.008}_{-0.008}$ \\
$q_{2,\mathrm{TESS}}$ & $\mathcal{N}(0.3835, 0.0078)$ & $0.384^{+0.007}_{-0.006}$ \\
$q_{1,\mathrm{CHEOPS}}$ & $\mathcal{N}(0.445, 0.0116)$ & $0.444^{+0.009}_{-0.010}$ \\
$q_{2,\mathrm{CHEOPS}}$ & $\mathcal{N}(0.4095, 0.0077)$ & $0.411^{+0.006}_{-0.006}$ \\
$q_{1,\mathrm{Sloan}\ g'}$ & $\mathcal{N}(0.6306, 0.0201)$ & $0.629^{+0.016}_{-0.016}$ \\
$q_{2,\mathrm{Sloan}\ g'}$ & $\mathcal{N}(0.4525, 0.0108)$ & $0.454^{+0.009}_{-0.009}$ \\
$q_{1,\mathrm{Sloan}\ r'}$ & $\mathcal{N}(0.4053, 0.013)$ & $0.404^{+0.010}_{-0.011}$ \\
$q_{2,\mathrm{Sloan}\ r'}$ & $\mathcal{N}(0.3979, 0.0092)$ & $0.397^{+0.007}_{-0.008}$ \\
$q_{1,\mathrm{Sloan}\ i'}$ & $\mathcal{N}(0.3079, 0.0115)$ & $0.300^{+0.007}_{-0.008}$ \\
$q_{2,\mathrm{Sloan}\ i'}$ & $\mathcal{N}(0.386, 0.0102)$ & $0.385^{+0.008}_{-0.009}$ \\
$q_{1,\mathrm{Pan\mbox{-}STARRS}\ z_{\mathrm{s}}}$ & $\mathcal{N}(0.2499, 0.0078)$ & $0.248^{+0.006}_{-0.006}$ \\
$q_{2,\mathrm{Pan\mbox{-}STARRS}\ z_{\mathrm{s}}}$ & $\mathcal{N}(0.376, 0.0082)$ & $0.375^{+0.006}_{-0.007}$ \\
\hline
\end{tabular}
\begin{tablenotes}
\item
Notes: $^{(a)}$For the priors, $\mathcal{N}$($\mu$, $\sigma$) indicates a normal distribution with mean $\mu$ and standard deviation $\sigma$, $\mathcal{U}$(\textit{a}, \textit{b}) a uniform distribution between \textit{a} and \textit{b}, and $\mathcal{F}$(\textit{a}) a parameter fixed to value \textit{a}.
$^{(b)}$The posterior estimate indicates the median value and then error bars for the 68 $\%$ credibility intervals. $^{(c)}$The orbit of TOI-1699A b is fixed to circular in the final fit. The $3\sigma$ upper limit is derived from a separate fit with free eccentricity (see Sect.~\ref{sec:join}).
\end{tablenotes}
\label{tab:posteriors}
\end{center}
\end{table*}
\clearpage
\section{Interior modeling for TOI-1699A b with \plaNETic}
\begin{table*}[!htbp]
\renewcommand{\arraystretch}{1.5}
\caption{Results of the internal structure modeling for TOI-1699A b.}
\centering
\begin{tabular}{lcccccc}
\hline \hline
Water prior &              \multicolumn{3}{c}{Water-rich} & \multicolumn{3}{c}{Water-poor} \\
Si/Mg/Fe prior &           Stellar (A1) &       Iron-enriched (A2) &      Free (A3) &
                           Stellar (B1) &       Iron-enriched (B2) &      Free (B3) \\
\hline
w$_\textrm{core}$ [\%] &        $11_{-8}^{+8}$ &    $15_{-11}^{+14}$ &    $12_{-9}^{+16}$ &
                           $16_{-11}^{+11}$ &    $21_{-15}^{+19}$ &    $17_{-13}^{+21}$ \\
w$_\textrm{mantle}$ [\%] &      $58_{-13}^{+16}$ &    $52_{-15}^{+19}$ &    $54_{-16}^{+19}$ &
                           $82_{-11}^{+11}$ &    $76_{-19}^{+15}$ &    $80_{-22}^{+13}$ \\
w$_\textrm{envelope}$ [\%] &    $30_{-18}^{+15}$ &    $30_{-18}^{+15}$ &    $30_{-18}^{+15}$ &
                           $1.99_{-0.28}^{+0.27}$ &    $2.54_{-0.62}^{+0.54}$ &    $2.26_{-0.82}^{+0.89}$ \\
\noalign{\medskip}
Z$_\textrm{envelope}$ [\%] &        $83.3_{-17.7}^{+6.6}$ &    $81.6_{-19.1}^{+7.2}$ &    $83.2_{-18.0}^{+7.0}$ &
                           $0.5_{-0.2}^{+0.2}$ &    $0.5_{-0.2}^{+0.2}$ &    $0.5_{-0.2}^{+0.2}$ \\
\noalign{\medskip}
x$_\textrm{Fe,core}$ [\%] &     $90.3_{-6.4}^{+6.5}$ &    $90.3_{-6.4}^{+6.5}$ &    $90.3_{-6.3}^{+6.5}$ &
                           $90.3_{-6.4}^{+6.6}$ &    $90.4_{-6.4}^{+6.5}$ &    $90.4_{-6.4}^{+6.5}$ \\
x$_\textrm{S,core}$ [\%] &      $9.7_{-6.5}^{+6.4}$ &    $9.7_{-6.5}^{+6.4}$ &    $9.7_{-6.5}^{+6.3}$ &
                           $9.7_{-6.6}^{+6.4}$ &    $9.6_{-6.5}^{+6.4}$ &    $9.6_{-6.5}^{+6.4}$ \\
\noalign{\medskip}
x$_\textrm{Si,mantle}$ [\%] &   $43_{-6}^{+7}$ &    $38_{-10}^{+9}$ &    $33_{-24}^{+31}$ &
                           $43_{-6}^{+7}$ &    $38_{-10}^{+9}$ &    $35_{-24}^{+30}$ \\
x$_\textrm{Mg,mantle}$ [\%] &   $41_{-6}^{+6}$ &    $36_{-9}^{+9}$ &    $38_{-26}^{+33}$ &
                           $41_{-6}^{+6}$ &    $36_{-9}^{+9}$ &    $37_{-25}^{+31}$ \\
x$_\textrm{Fe,mantle}$ [\%] &   $16_{-11}^{+9}$ &    $25_{-17}^{+19}$ &    $19_{-15}^{+24}$ &
                           $17_{-11}^{+9}$ &    $25_{-17}^{+19}$ &    $20_{-15}^{+24}$ \\
\hline
\end{tabular}
\label{tab:internal_structure_results_b}
\end{table*}
\renewcommand{\arraystretch}{1.0}

\end{document}